arXiv:1802.00793v1 [econ.EM] 2 Feb 2018

# Structural analysis with mixed-frequency data: A MIDAS-SVAR model of US capital flows

Emanuele Bacchiocchi* Andrea Bastianin†

Alessandro Missale‡ Eduardo Rossi§

1st February 2018

**Abstract**

We develop a new VAR model for structural analysis with mixed-frequency data. The MIDAS-SVAR model allows to identify structural dynamic links exploiting the information contained in variables sampled at different frequencies. It also provides a general framework to test homogeneous frequency-based representations versus mixed-frequency data models. A set of Monte Carlo experiments suggests that the test performs well both in terms of size and power. The MIDAS-SVAR is then used to study how monetary policy and financial market volatility impact on the dynamics of gross capital inflows to the US. While no relation is found when using standard quarterly data, exploiting the variability present in the series within the quarter shows that the effect of an interest rate shock is greater the longer the time lag between the month of the shock and the end of the quarter.

**Keywords:** Mixed frequency variables, capital flows, monetary policy, volatility shocks.

*University of Milan, Department of Economics, Management and Quantitative Methods, Via Conservatorio 7, 20122 Milan, Italy. Email: emanuele.bacchiocchi@unimi.it.

†University of Milan, Department of Economics, Management and Quantitative Methods, Via Conservatorio 7, 20122 Milan, Italy. Email: andrea.bastianin@unimi.it.

‡University of Milan, Department of Economics, Management and Quantitative Methods, Via Conservatorio 7, 20122 Milan, Italy. Email: alessandro.missale@unimi.it.

§University of Pavia, Department of Economics and Management, Via San Felice 5, 27100, Pavia, Italy. Email: eduardo.rossi@unipv.it.

# 1 Introduction

A standard assumption in the empirical macroeconometric literature is that agents make their decisions at fixed intervals of time. For practical reasons such time intervals usually coincide with the sampling frequency of the data used in the analysis. But, Christiano and Eichenbaum (1987) show that a specification error, that they term "temporal aggregation bias", affects both parameter estimates and hypothesis testing when economic agents make decisions at time intervals that are finer and hence do not match the sampling frequency of the data. Bayar (2014) analyzes the impact of temporal aggregation when estimating Taylor-type monetary policy rules. Using interest rates, that are averaged at quarterly frequency to match other macroeconomic variables, leads to overestimating the interest rate smoothing parameter, and thus to a biased assessment of the persistence of monetary policy changes. Foroni and Marcellino (2014) show that the temporal aggregation of dynamic stochastic general equilibrium (DSGE) models from monthly to quarterly frequency significantly distorts the estimated responses to a monetary policy shock.

A partial solution to the temporal aggregation bias is provided by econometric methodologies developed to analyze variables measured at different sampling frequencies, the so-called Mixed Data Sampling regression models (MIDAS).[1] However, the MIDAS literature has been mainly concerned with reduced-form univariate time series models, focusing on the potential information embedded in high frequency data to better forecast low frequency variables.[2] Only recently, authors have started dealing with multivariate models for mixed frequency data. This literature has developed along two main lines of research. A first approach, pioneered by Zadrozny (1990), assumes that there is a high-frequency latent process for which only low-frequency observations are available. State-space representation combined with Bayesian or classical estimation are used to match the latent process with the observed low-frequency data.[3] A second approach is that of treating all variables at different frequencies as a "stacked skip-sampled" process; the relations between variables are then jointly analyzed in a way that is much closer to the tradition of Vector Autoregressive (VAR) models (see Ghysels, 2016). Since latent factors and latent shocks are absent, models in this strand of the literature are straightforward multivariate extensions of the univariate MIDAS regression model.[4]

Models for mixed frequency data can uncover dynamic or structural relationships that would otherwise be hidden by the temporal aggregation of data available at different sampling frequencies, such as macroeconomic flow variables and financial variables. The mixed-frequency problem emerges in the analysis of monetary policy and financial markets. For instance daily data are often used to identify monetary policy shocks at monthly frequencies or to study

[1]See, for a survey, Foroni, Ghysels, and Marcellino (2014).

[2]See, among many others, Clements and Galvao (2008, 2009).

[3]See, among others, Mariano and Murasawa (2003); Giannone, Reichlin, and Small (2008); Arouba, Diebold, and Scotti (2009); Eraker, Chiu, Foerster, Kim, and Seoane (2015); Schorfheide and Song (2015).

[4]See Andreou, Ghysels, and Kourtellos (2010).

relationship between monetary policy and financial markets.[5]

The temporal aggregation problem naturally arises in the growing literature on the global financial cycle that investigates the determinants of international capital flows as data on the latter are available only at a quarterly frequency. For instance, the use of a quarterly structural VAR model by Bruno and Shin (2015) may explain the short-lived and not always statistically significant relationships between US interest rates, the Market Volatility index (VIX), the real effective exchange rate and cross-border bank capital flows. A natural question is whether these empirical findings can somehow be explained by the existence of a "temporal aggregation bias", arising when transforming high frequency variables, such as interest rates and VIX, to match lower frequency variables, such as gross capital flows.

In this paper, unlike in most existing work on mixed frequency models which focuses on forecasting, we examine the response of macroeconomic and financial variables to structural shocks identified relying on variables observed at different frequencies. In this sense, this paper complements the analysis by Foroni and Marcellino (2014) and Christensen, Posch, and van der Wel (2016) who emphasize the role of mixed frequency variables in DSGE models. To this end, following Ghysels (2016), we propose a mixed frequency VAR model, the MIDAS-VAR, that can be seen as a multivariate specification that encompasses both the unrestricted-MIDAS (U-MIDAS) model by Foroni, Marcellino, and Schumacher (2014) and the reverse unrestricted MIDAS (RU-MIDAS) model by Foroni, Guérin, and Marcellino (2015). We also present and discuss the structural representation of the MIDAS-VAR model and provide conditions for the identification of its structural parameters. Our relatively simple parametrization allows the standard mix of OLS and maximum likelihood (ML) techniques to provide consistent estimates for the reduced- and structural-form parameters, respectively. Furthermore, we present a formal test procedure to evaluate the benefits of using a structural MIDAS-VAR model (MIDAS-SVAR, henceforth) with respect to traditional low frequency structural VAR models (SVAR, henceforth). The performance of this test is analyzed in a set of Monte Carlo experiments.

Second, we investigate the relationships between US capital inflows, monetary policy and financial market volatility, as measured by the VIX, using a MIDAS-SVAR for a mixture of quarterly data on gross capital inflows and monthly data on the policy interest rate and the VIX. By using both sampling frequencies. We find that a monetary policy shock has a different impact on capital inflows depending on the month it happens within the quarter. The effect is positive and statistically significant on impact when the interest-rate shock occurs in the first month of the quarter, positive but statistically significant only after two quarters when the shock occurs in the second month, while it becomes negative when the shock takes place in the third month, i.e. at the end of the quarter. This can be explained by the high observed persistence of shocks to the Federal Funds rate. A persistent increase in the policy rate occurring at the beginning of the quarter can have a sizable effect on capital flows over the quarter because such flows accumulate over the entire three-month period. Moreover, we find that a shock to the VIX leads to an immediate and statistically significant reduction of capital flows, independently of the month it happens within the quarter, with the strongest impact observed when the shock takes place in the second month of the quarter.

[5]See, e.g., Faust, Rogers, Swanson, and Wright (2003) and Cochrane and Piazzesi (2002) for identification using daily data and Rigobon and Sack (2003), Rigobon and Sack (2004) and Bekaert, Hoerova, and Lo Duca (2013) for an analysis of monetary policy and markets.

The rest of the paper is organized as follows: In Section 2 we introduce the MIDAS-VAR and MIDAS-SVAR models, discuss the identification conditions of the structural parameters and provide some results on the relationships between these models and the traditional VAR and SVAR models. In Section 3 we present a test strategy to statistically evaluate the performance of a MIDAS-VAR when compared to a standard VAR. Section 4 is dedicated to the empirical analysis, with a discussion on the identification of the structural shocks and on the estimated Impulse Responses (IRFs) and Forecast Error Variance Decompositions (FEVDs). Section 5 presents a rich set of robustness checks. Section 6 concludes. Additional technical details and empirical results are confined in a Technical Supplement associated with the paper.

## 2 The MIDAS-SVAR model: representation and identification

In this section we introduce a multivariate model for investigating the structural relationships between variables observed at different frequencies, as is the case for gross capital inflows, interest rates and the Market Volatility index (VIX).

### 2.1 Representation

Consider two vectors of variables $x_L$ and $x_H$ containing the $n_L$ low-frequency and $n_H$ high-frequency variables, respectively, where $x_H$ are sampled $m$ times more often than $x_L$. The MIDAS-VAR model, when considering quarterly and monthly series, i.e. $m = 3$, can be written as[6]

$$\begin{pmatrix} x_H(t,1) \\ x_H(t,2) \\ x_H(t,3) \\ x_L(t) \end{pmatrix} = \sum_{i=1}^{p} \begin{pmatrix} A_{11}^i & A_{12}^i & A_{13}^i & A_1^i \\ A_{21}^i & A_{22}^i & A_{23}^i & A_2^i \\ A_{31}^i & A_{32}^i & A_{33}^i & A_3^i \\ A_{L1}^i & A_{L2}^i & A_{L3}^i & A_L^i \end{pmatrix} \begin{pmatrix} x_H(t-i,1) \\ x_H(t-i,2) \\ x_H(t-i,3) \\ x_L(t-i) \end{pmatrix} + \begin{pmatrix} u_H(t,1) \\ u_H(t,2) \\ u_H(t,3) \\ u_L(t) \end{pmatrix} \tag{1}$$

where the time index $t$ refers to the low-frequency variables, while for the high-frequency variables the couple $(t, j)$ indicates the month $j$ of observation within the quarter $t$, and where $p$ represents the order of the MIDAS-VAR. Denoting the vector of observable variables (sampled at different frequency) as $\tilde{x}(t)$, with dimension $\tilde{n} \times 1$, being $\tilde{n} = n_L + mn_H$, allows us to provide

[6] Deterministic components, such as constant term, intervention dummies, time trend, are omitted for simplicity but can be managed as in the traditional VAR literature.

the following more compact notation equivalent to the one used for traditional VAR models:

$$A\left(L\right)\tilde{x}\left(t\right)=\tilde{u}\left(t\right) \tag{2}$$

where $L$ denotes the low-frequency lag operator, i.e. $Lx_L\left(t\right)=x_L(t-1)$ and $Lx_H\left(t,j\right)=x_H(t-1,j)$, and $A(L)=I_{\tilde{n}}-\sum_{i=1}^{p}A_iL^i$.

The representation in Eqs. (1)-(2), being a function of the past values of the observable variables, though at different frequencies, can be seen as the reduced form of the model. However, looking, for example, at the (multivariate) equation for the vector of variables $x_H\left(t,2\right)$, it is immediately clear that this does not depend on its first natural lag, $x_H\left(t,1\right)$. The correct specification of the dynamics of the process does not prevent the error term $u\left(t,2\right)$ to be correlated with $u\left(t,1\right)$. More generally, the covariance matrix of the error terms is defined as

$$\Sigma_{\tilde{u}}=\begin{pmatrix}\Sigma_{11} & & & \\ \Sigma_{21} & \Sigma_{22} & & \\ \Sigma_{31} & \Sigma_{32} & \Sigma_{33} & \\ \Sigma_{L1} & \Sigma_{L2} & \Sigma_{L3} & \Sigma_{L}\end{pmatrix} \tag{3}$$

and none of the blocks is supposed to be zero.

The covariance matrix $\Sigma_{\tilde{u}}$, thus, contains all contemporaneous relations among the high-frequency variables ($\Sigma_{11}$, $\Sigma_{22}$ and $\Sigma_{33}$), among the low-frequency variables ($\Sigma_L$), the within-quarter relations between low- and high-frequency variables ($\Sigma_{L1}$, $\Sigma_{L2}$ and $\Sigma_{L3}$) and some further dynamic relations among the high-frequency variables ($\Sigma_{21}$, $\Sigma_{31}$ and $\Sigma_{32}$).

## 2.2 Identification of structural relationships

Ghysels (2016) discusses possible implementations of structural mixed frequency VAR models and, in particular, proposes the distinction between real-time predictions and policy response functions. As we are interested in the latter, i.e. in understanding the structural relationships among the variables, all the relations discussed at the end of the previous section, hidden in the covariance matrix $\Sigma_{\tilde{u}}$, must be explicitly identified.

As common in the literature, we perform policy analysis through impulse response functions (IRFs) describing the dynamic transmission of uncorrelated structural shocks among the variables. Based on the MIDAS-VAR model proposed in Eq. (1), under the assumption of stationarity, the IRFs can be easily obtained through the MIDAS-VMA representation

$$\begin{aligned}\tilde{x}(t) &= \left(I_{\tilde{n}}-\sum_{i=1}^{p}A_iL^i\right)^{-1}\tilde{u}(t)\\ &= \sum_{k=0}^{\infty}C_k\tilde{u}(t-k)\equiv C(L)\tilde{u}(t)\end{aligned}$$

where $I_{\tilde{n}} = A(L)C(L)$. More specifically, the IRFs generally refer to the $(\tilde{n} \times 1)$ vector of latent uncorrelated structural shocks defined as

$$A\tilde{u}(t) = B\tilde{\varepsilon}(t) \qquad \text{with} \qquad \tilde{\varepsilon}_t \sim (0, I_{\tilde{n}}) \tag{4}$$

generating a set of non-linear relationships $\Sigma_{\tilde{u}} = A^{-1}B\Sigma_{\tilde{\varepsilon}}B'A^{-1\prime}$, that link reduced-form moments with the structural parameters $A$ and $B$, with $A$ and $B$ non-singular $\tilde{n} \times \tilde{n}$ matrices, and the covariance matrix of the structural shocks $\Sigma_{\tilde{\varepsilon}} = I_{\tilde{n}}$ as in Eq. (4).

**Remark: MIDAS-SVAR, U-MIDAS and RU-MIDAS.** Foroni, Marcellino, and Schumacher (2014) provide formal derivation of single equation Unrestricted MIDAS (U-MIDAS) models where high-frequency variables are exploited to improve the forecast of low-frequency variables. Specifically they obtain an *exact* U-MIDAS representation where the error term enters with a moving average structure (see, e.g., Marcellino (1999) and the references therein). However, as the parameters of such a structure cannot be exactly determined, they provide an *approximate* version where enough dynamics is included in order to make the residuals as close as possible to the realization of a white noise process. Similarly, Foroni, Guérin, and Marcellino (2015) derive the *exact* and *approximate* Reverse Unrestricted MIDAS (RU-MIDAS) models where low-frequency variables are incorporated in models for predicting high frequency variables. The stacked presentation of the MIDAS-VAR model adopted in this section allows to handle the set of equations for the low-frequency variables, at the bottom of the model, exactly as the *approximate* U-MIDAS model. Furthermore, as it will be clear in the empirical application described in the next sections, the $A$ matrix in Eq. (4) makes the equations for the high-frequency variables exactly equivalent to the *approximate* RU-MIDAS models. The $A$ matrix, in fact, helps to include the dynamics of the high-frequency variables naturally missing in the formulation of the reduced-form specification in Eq. (1).■

The AB-MIDAS-SVAR model, or more simply, the MIDAS-SVAR model described in Eqs. (1) and (4) provides a very general framework for investigating the contemporaneous relations between observable high- and low-frequency variables from one side, and between high- and low-frequency structural shocks from the other, captured by the $A$ and $B$ matrices, respectively. This specification, deeply investigated in Amisano and Giannini (1997) for the traditional SVAR models, is more general than the one proposed in Ghysels (2016) where the only considered source of structural relationships is confined to the $A$ matrix, restricting the $B$ matrix to be diagonal. While many empirical applications of SVAR models regarding the transmission of the monetary policy focus on the $B$ matrix (fixing $A$ equal to the identity matrix), Bernanke (1986), Blanchard (1989) and Blanchard and Perotti (2002) provide examples in which both simultaneous relations are possible.

Clearly, the identification of $A$ and $B$, and, as a consequence, of the latent structural shocks, is subject to restrictions on the parameters. In fact, following the definition of the AB-MIDAS-SVAR model provided in Eq. (4), there are $2\tilde{n}^2$ parameters to be estimated from the $\tilde{n}(\tilde{n}+1)/2$ empirical moments contained in the $\Sigma_u$. The following proposition, from Lütkepohl (2006), provides a necessary and sufficient condition for the identification of the two matrices $A$ and $B$ when subjected to the linear restrictions given by

$$S_A vec\, A = s_A \qquad \text{and} \qquad S_B vec\, B = s_B \tag{5}$$

for some $S_A$, $s_A$, $S_B$ and $s_B$ known matrices.

**Proposition 1.** *(Local identification of the AB-model)*
*Consider the AB-MIDAS-SVAR model reported in Eqs. (1) and (4), subject to the restrictions in Eq. (5). For a given covariance matrix of the error terms $\Sigma_{\tilde{u}}$, the parameters in $A$ and $B$ are locally identifiable if and only if*

$$rank \begin{pmatrix} -2D_{\tilde{n}}^{+}\left(\Sigma_{\tilde{u}} \otimes A^{-1}\right) & 2D_{\tilde{n}}^{+}\left(A^{-1}B \otimes A^{-1}\right) \\ S_A & 0 \\ 0 & S_B \end{pmatrix} = 2\tilde{n}^2. \tag{6}$$

*Proof.* The Proposition can be proved by calculating the Jacobian and applying the results in Rothenberg (1971).[7] □

Ghysels (2016) discusses the importance of the triangular Cholesky factorization in the mixed frequency VAR framework, where the high-frequency variables have a natural order for intra-$t$ timing of shocks. Ghysels focuses on the potential source of information coming from the high-frequency variables in explaining the dynamics of the low-frequency ones, whereas the AB-MIDAS-SVAR framework offers a very flexible tool to examine the interaction of low- and high-frequency variables, and provides a better understanding of macroeconomic variables fluctuations.

# 3 Testing the equivalence of MIDAS-SVAR and SVAR

Ghysels (2016) discusses the asymptotic properties of misspecified VAR model estimators, where the misspecification arises from a wrong selection of the sampling frequency of the variables. The analysis proceeds by considering a high-frequency VAR model (characterized by latent processes if the actual series are observed at a lower frequency only), and compares it to a low-frequency model obtained by certain aggregation schemes. In particular, his Proposition 5.1 states that the asymptotic impact of misspecification is a function of the aggregation scheme. In what follows, we restrict the analysis to the two types of VAR models used in the empirical analysis: A low-frequency quarterly VAR and a monthly-quarterly MIDAS-VAR. The aim is to provide a testing strategy to verify whether aggregating the data and passing to a low-frequency VAR generates a substantial loss of information that might invalidate the analysis. We first focus on the dynamics of the VAR, and then move to the structural part of the model.

[7] See Lütkepohl (2006), page 365, for details.

### 3.1 Matching the dynamics

Let us assume that $\tilde{x}(t)$ has a MIDAS-VAR representation. The following proposition establishes the equivalence in conditional mean between MIDAS-VAR for $\tilde{x}(t)$ and VAR representation for homogeneous sampling frequency variables in $\ddot{x}(t)$, obtained as

$$\ddot{A}(L)\ddot{x}(t) = \ddot{u}_t. \tag{7}$$

**Proposition 2.** *Let $\ddot{x}(t) = G\tilde{x}(t)$ be a vector containing variables sampled at the same frequency, the MIDAS-VAR(p) representation for $\tilde{x}(t)$ is equivalent in mean to the VAR(p) representation for $\ddot{x}(t)$ if*

$$GA_i\tilde{x}(t-i) = \ddot{A}_i\ddot{x}(t-i), \quad i = 1, 2, \ldots, p. \tag{8}$$

*Proof.* See Appendix A.1. □

The equivalence, i.e. the condition in Eq. (8), holds if the matrices $A_i, i = 1, \ldots, p$ are appropriately restricted. To illustrate this, let us start from a MIDAS-VAR model in which we have monthly and quarterly variables, with just one quarterly-lagged value:

$$\begin{pmatrix} x_H^1(t,1) \\ x_H^1(t,2) \\ x_H^1(t,3) \\ x_L^2(t) \end{pmatrix} = \begin{pmatrix} A_{11}^1 & A_{12}^1 & A_{13}^1 & A_1^1 \\ A_{21}^1 & A_{22}^1 & A_{23}^1 & A_2^1 \\ A_{31}^1 & A_{32}^1 & A_{33}^1 & A_3^1 \\ A_{L1}^1 & A_{L2}^1 & A_{L3}^1 & A_L^1 \end{pmatrix} \begin{pmatrix} x_H^1(t-1,1) \\ x_H^1(t-1,2) \\ x_H^1(t-1,3) \\ x_L^2(t-1) \end{pmatrix} + \begin{pmatrix} u_H^1(t,1) \\ u_H^1(t,2) \\ u_H^1(t,3) \\ u_L^2(t) \end{pmatrix} \tag{9}$$

where $x_H^1$ collects the monthly series while $x_L^2$ the quarterly ones. This specification should be compared to a traditional VAR in which both groups of variables are observed at the same quarterly frequency, i.e.

$$\begin{pmatrix} \ddot{x}_L^1(t) \\ \ddot{x}_L^2(t) \end{pmatrix} = \begin{pmatrix} \ddot{A}_{11}^1 & \ddot{A}_{12}^1 \\ \ddot{A}_{21}^1 & \ddot{A}_{22}^1 \end{pmatrix} \begin{pmatrix} \ddot{x}_L^1(t-1) \\ \ddot{x}_L^2(t-1) \end{pmatrix} + \begin{pmatrix} \ddot{u}_L^1(t) \\ \ddot{u}_L^2(t) \end{pmatrix} \tag{10}$$

where $\ddot{x}_L^2 = x_L^2$.

The comparison between the two specifications depends on the form of the time aggregation used to transform $x_H^1(t,1)$, $x_H^1(t,2)$ , $x_H^1(t,3)$ into $\ddot{x}_L^1(t)$. The mapping from the MIDAS-VAR to the VAR model reduces to consider a selection matrix $G$ accounting for the function used to aggregate the high frequency variables. If the aggregation scheme consists in taking the first observation of the quarter only, the $G$ matrix becomes:

$$G = \begin{pmatrix} I_{n_H} & 0 & 0 & 0 \\ 0 & 0 & 0 & I_{n_L} \end{pmatrix} \tag{11}$$

and the associated null hypothesis for testing the equivalence between the two specifications (without exogenous variables, for simplicity) reduces to[8]

$$\begin{aligned} H_0^1 &: A_{12}^1 = A_{13}^1 = 0 \\ H_0^2 &: A_{L2}^1 = A_{L3}^1 = 0. \end{aligned} \tag{12}$$

A natural way to evaluate the benefits of relying on mixed-frequency data instead of temporally aggregated data is to implement a Wald- or LR-type test for the joint null hypothesis $H_0^1$ and $H_0^2$, against the alternative that at least one of the two is not supported by the data. If the aim of the analysis is to use the dynamics of the model to forecast the future values of the endogenous variables, under the null hypothesis in Eq. (12), mixing monthly and quarterly observations does not provide any gain. Under the alternative, the information provided by mixed-frequency data is statistically relevant and useful to obtain more accurate forecasts.

From Proposition 2 we can also obtain the IRFs of $\bar{x}(t)$ w.r.t. the shocks in the MIDAS-VAR. Under restriction in Eq. (8), we have

$$\ddot{x}(t) = \ddot{A}(L)^{-1} G\tilde{u}(t) = \ddot{C}(L) G\tilde{u}(t)$$

## 3.2 Matching the structural relationships

In the following proposition we define the restrictions for the equivalence between the structural relations in the MIDAS-SVAR model, Eq.(1) and Eq.(4), and those in the SVAR model for the homogeneous sampling frequency variables, $\ddot{x}(t)$, $\bar{A}\bar{u}(t) = \bar{B}\bar{\varepsilon}(t)$.

**Proposition 3.** *Let $\ddot{x}(t) = G\tilde{x}(t)$ be a vector containing variables sampled at the same frequency, the MIDAS-SVAR representation for $\tilde{x}(t)$ is equivalent to the SVAR representation for*

[8] Details on the derivation of the null hypothesis are provided in Appendix A.

$\ddot{x}(t)$ *if*

$$GA\tilde{u}(t) = \ddot{A}\ddot{u}(t)$$
$$GB\tilde{\varepsilon}(t) = \ddot{B}\ddot{\varepsilon}(t)$$

*Proof.* See Appendix A.1. □

As an example, the MIDAS-SVAR model considered in Section 2.2, when the frequency of the data is both monthly and quarterly, can be written as

$$\underbrace{\begin{pmatrix} A_{11} & A_{12} & A_{13} & A_1 \\ A_{21} & A_{22} & A_{23} & A_2 \\ A_{31} & A_{32} & A_{33} & A_3 \\ A_{L1} & A_{L2} & A_{L3} & A_L \end{pmatrix}}_{A} \underbrace{\begin{pmatrix} u_H^1(t,1) \\ u_H^1(t,2) \\ u_H^1(t,3) \\ u_L^2(t) \end{pmatrix}}_{\tilde{u}(t)} = \underbrace{\begin{pmatrix} B_{11} & B_{12} & B_{13} & B_1 \\ B_{21} & B_{22} & B_{23} & B_2 \\ B_{31} & B_{32} & B_{33} & B_3 \\ B_{L1} & B_{L2} & B_{L3} & B_L \end{pmatrix}}_{B} \underbrace{\begin{pmatrix} \varepsilon_H^1(t-1,1) \\ \varepsilon_H^1(t-1,2) \\ \varepsilon_H^1(t-1,3) \\ \varepsilon_L^2(t-1) \end{pmatrix}}_{\tilde{\varepsilon}(t)} \tag{13}$$

with $\tilde{u}(t)$ and $\tilde{\varepsilon}(t)$ defined as in Section 2.2 and where the elements in $A$ and $B$ must be restricted in order to fulfill the rank condition in Proposition 1. In the SVAR model with aggregated quarterly data, the specification of the structural relationships is given by

$$\begin{pmatrix} \ddot{A}_{11} & \ddot{A}_{12} \\ \ddot{A}_{21} & \ddot{A}_{22} \end{pmatrix} \begin{pmatrix} \ddot{u}_L^1(t) \\ \ddot{u}_L^2(t) \end{pmatrix} = \begin{pmatrix} \ddot{B}_{11} & \ddot{B}_{12} \\ \ddot{B}_{21} & \ddot{B}_{22} \end{pmatrix} \begin{pmatrix} \ddot{\varepsilon}_L^1(t) \\ \ddot{\varepsilon}_L^2(t) \end{pmatrix} \tag{14}$$

where $\left(\ddot{u}_L^{1\prime}(t), \ddot{u}_L^{2\prime}(t)\right)'$ are the residuals of the quarterly VAR as in Eq. (10) and $\left(\ddot{\varepsilon}_L^{1\prime}(t), \ddot{\varepsilon}_L^{2\prime}(t)\right)'$ are the quarterly structural shocks.

If quarterly observations are obtained by simply taking the first month of the quarter, as in Section 3.1, the two specifications in Eq.(13) and (14) can be compared by using the selection matrix $G$ introduced in Eq. (11). Premultiplying Eq. (13) by $G$, allows us to show the relation between the MIDAS-SVAR and the quarterly SVAR. In particular, if the following relations

$$\begin{aligned} H_0^1 &: A_{12} = A_{13} = 0 \\ H_0^2 &: A_{L2} = A_{L3} = 0 \\ H_0^3 &: B_{12} = B_{13} = 0 \\ H_0^4 &: B_{L2} = B_{L3} = 0 \end{aligned} \tag{15}$$

hold, the two specifications are statistically equivalent, and using monthly data does not add useful information to identify the structural shocks.

If the aim of the analysis is to identify the structural shocks and to understand their transmission mechanisms, a statistical test can be implemented to verify whether a (monthly-quarterly) MIDAS-SVAR has to be preferred to a traditional (quarterly) SVAR. The test consists in jointly testing the null hypotheses in Eqs. (12)-(15) that can be implemented through a standard LR- or Wald-type test strategy. As an example, the implementation of a LR-type test reduces to calculate the log-likelihood of the unrestricted model ($l^u$) and that of the restricted one ($l^r$) according to both the identifying restrictions and those in Eqs. (12)-(15). The test statistic $LR = -2\left(l^r - l^u\right)$, as well-known, is asymptotically distributed as a $\chi^2$ with the number of degrees of freedom (dof) given by the order of the over-identification. If the null hypothesis is rejected, aggregating the data loses substantial information that instead is important in the identification of the structural shocks.

### 3.3 A Monte Carlo evaluation of the LR-test performance

We are interested in evaluating the performance of the test of hypothesis introduced in the previous section to statistically check the equivalence of the MIDAS-VAR model to the traditional one. While the asymptotic distribution of the test statistic is standard, the behavior in small samples can be problematic due to the large number of restrictions imposed. We provide results for different data generating processes (DGPs), focusing on the reduced-form of the model. The number of observation is $T = 109$ and is the same for all experiments (it corresponds to the number of quarterly observations in the empirical analysis presented below). We first provide some simulations for a small-scale MIDAS-VAR model, and then we move to a larger setting as the one applied in the empirical analysis. All the simulations are evaluated through the *P-value plot* and the *size-power curve* introduced in Davidson and MacKinnon (1998); for each simulation we provide results obtained with 5,000 replications.

#### 3.3.1 Small-scale MIDAS-VAR

The first data generating process we consider (indicated as **Model 1-$H_1$**) is a MIDAS-VAR with just one quarterly variable and one monthly variable ($n_L = 1$ and $n_H = 1$). In particular, the DGP is obtained through the estimation of a MIDAS-VAR, with one single (quarterly) lag, where the high-frequency variable, $x_H$, is the short-term interest rate and the low-frequency one, $x_L$, is the ratio of gross capital inflows to GDP that will be described in the empirical analysis in the following section. This makes the simulated model a small version of that used in the empirical analysis. Following the notation in Eq. (1), the DGP is given by the following parameters

$$A_1 = \begin{pmatrix} -0.066 & -0.344 & 1.398 & 0.003 \\ -0.319 & -0.352 & 1.641 & 0.003 \\ -0.505 & -0.248 & 1.727 & 0.003 \\ -4.005 & 0.340 & 3.276 & 0.524 \end{pmatrix}, \quad \Sigma_{\tilde{u}} = \begin{pmatrix} 0.029 & 0.051 & 0.065 & 0.385 \\ 0.051 & 0.119 & 0.153 & 0.873 \\ 0.065 & 0.153 & 0.219 & 0.935 \\ 0.385 & 0.873 & 0.935 & 94.231 \end{pmatrix} \tag{16}$$

where the vector of constant terms is not reported for simplicity. The parameters in Eq. (16) generate the data under the (alternative) hypothesis of a MIDAS-VAR where the different nature of the variables clearly matters. In order to evaluate both the size and the power of the test developed in Section 3.1, we also generate the data using a DGP obtained by the estimation of a quarterly VAR model with the same two variables as before but with the interest rate observed at the first month of the quarter (**Model 1-**$H_0$). In Figure 1 we report a graphical evaluation of both the size and the power of the test through the P-value plot (left panel) and the size-power curve (right panel). The P-value plot is the simple empirical distribution function (EDF) of the p-value of the test against a set of points in the interval (0,1).[9] If the distribution of the test used to calculate the p-value is correct, the P-value plot should be close to the 45° line. Figure 1, left panel, provides evidence on the actual size of the test when the DGP is **Model 1-**$H_0$ and the null hypothesis is as in Eq. (12), where, given the simplicity of the model, all $A^1_{12}$, $A^1_{13}$, $A^1_{L2}$ and $A^1_{L3}$ are scalars. Thus, the LR test statistic is asymptotically distributed as a $\chi^2$ with 4 dof. As expected, given the reduced number of parameters, the actual size of the test practically corresponds to the theoretical one, at all significance levels. Figure 1, right panel, instead, provides evidence on the power of the test and plots the EDF of the p-value when the DGP is given by **Model 1-**$H_1$ (Y-axis) against the EDF of the p-values when the DGP is given by **Model 1-**$H_0$ (X-axis). The power of the test, thus, is plotted against the true size, instead of the nominal one. From Figure 1, right panel, it is immediately evident that the null hypothesis is rejected at all critical levels, providing strong evidence that **Model 1-**$H_1$ is too far away from the null hypothesis, and that the high-frequency nature of one of the variables contains information that cannot be neglected in the analysis.

To analyze the power of the test when the null and the alternative hypothesis are much closer than in the previous experiment, we generate two additional datasets (**Model 2-**$H_1$ and **Model 3-**$H_1$) using the same covariance matrix $\Sigma_{\tilde{u}}$, but with the following parameters describing the dynamics of the MIDAS-VARs:

$$\underset{\textbf{Model 2-}H_1}{A_1 = \begin{pmatrix} 0.3 & 0.1 & 0.1 & 0 \\ 0.1 & 0.2 & 0.3 & 0 \\ 0.1 & 0.1 & 0.4 & 0 \\ 0 & 0.1 & 0.1 & 0.5 \end{pmatrix}} \qquad \underset{\textbf{Model 3-}H_1.}{A_1 = \begin{pmatrix} 0.3 & 0.03 & 0.08 & 0 \\ 0.1 & 0.2 & 0.3 & 0 \\ 0.1 & 0.1 & 0.4 & 0 \\ 0 & 0.02 & 0.01 & 0.5 \end{pmatrix}} \tag{17}$$

For both experiments, the corresponding data are generated when the null hypothesis, as the one described in the first experiment, is true (**Model 2-**$H_0$ and **Model 3-**$H_0$). Apart from the *sensible* parameters involved in the null hypothesis, i.e. $A^1_{12}$, $A^1_{13}$, $A^1_{L2}$ and $A^1_{L3}$, all the remaining parameters are the same in the last two specifications. The size performance of the test, reported in Figure 1, left panel, are thus identical for **Model 2-**$H_0$ and **Model 3-**$H_0$. As for the previous experiment, the actual size of the test practically coincides with the nominal one, for all confidence levels. Concerning the power, shown in Figure 1, right panel, the test performs differently for the two models. In particular, the test continues to perform quite well for **Model 2-**$H_1$, where the power is about $75-80\%$ for the usual 1% and 5% critical levels. For **Model 3-**$H_1$, instead, the power enormously decreases although remaining quite satisfactory if one thinks that **Model 3-**$H_1$ and **Model 3-**$H_0$ are very close to each other.

[9]The EDF is evaluated at $m = 215$ points as suggested by Davidson and MacKinnon (1998).

### 3.3.2 Medium-scale MIDAS-VAR

This section considers experiments on MIDAS-VAR models where the number of variables involved, and thus the number of restrictions, increase. As before, we consider three experiments, one based on a DGP obtained through real data, while the other two based on artificial sets of parameters.

The first DGP (**Model 1-**$H_1$) is obtained by estimating a MIDAS-VAR using actual data, with $n_H = 3$ high-frequency monthly variables and $n_L = 1$ quarterly variable with one lag. The monthly variables are the short-term interest rate, an indicator of market volatility and an indicator of the business cycle, while the quarterly one is, as in the previous section, the gross capital inflow relative to GDP. The variables are described in the empirical section below. The matrices of parameters are reported in Appendix B. Using the same set of variables, we estimate a MIDAS-VAR model with the restrictions set in Eq. (12). Then, using this new set of parameters we generate the new datasets in which the null hypothesis is true (**Model 1-**$H_0$). In these latter sets of experiments, the null hypothesis is equivalent to the one reported in Eq. (12) and also in Section 3.3.1, but now $A^1_{12}$ and $A^1_{13}$ are $(3 \times 3)$ matrices while $A^1_{L2}$ and $A^1_{L3}$ are $(1 \times 3)$ vectors. The asymptotic distribution of the test statistic is thus a $\chi^2$ with 24 dof.

Size and power comparisons are shown in Figure 2. Due to the large number of restrictions, the empirical size tends to over-reject the null; at the canonical 5% and 10% critical levels, the test rejects at 10.4% and 18.4%, respectively. Concerning the power of the test, reported in the right panel, against the true size of the test, we can always reject the null hypothesis when the data are generated through **Model 1-**$H_1$. Given that the DGP is generated by estimating a MIDAS-VAR on real data, this result emphasizes the importance of considering all the information contained in the data, instead of aggregating into a quarterly VAR.

As before, however, we propose two additional models (**Model 2-**$H_1$ and **Model 3-**$H_1$) generating data more closely related to the null hypothesis. For evaluating the size of the test and drawing the size-power curve, the models that generate the data when the null is true (**Model 2-**$H_0$ and **Model 3-**$H_0$) are obtained by imposing the zero restrictions introduced in Eq. (12). As the only different parameters in **Model 2-**$H_1$ and **Model 3-**$H_1$ are those involved in the null hypothesis, the two specifications of **Model 2-**$H_0$ and **Model 3-**$H_0$ are equivalent. Figure 2, left panel, evaluates the empirical size of the test, which is very similar to the one already obtained for **Model 1-**$H_0$. Figure 2, right panel, evaluates the power of the test for the two models. The power is extremely high for **Model 2-**$H_1$ while it decreases for **Model 3-**$H_1$, in which the choice of the parameters is extremely close to the specification under the null hypothesis (**Model 3-**$H_0$). Despite this reduction, the power continues to be quite high and much larger than the empirical size of the test.

# 4 A MIDAS-SVAR analysis of US capital inflows and monetary policy

In light of the methodology developed in the previous section, we shall present new results emphasizing the role played by the *natural* mixed frequency of the variables in detecting the effect of monetary policy, and financial market volatility, on US gross capital inflows.

We estimate a three-equation MIDAS-SVAR model for US gross capital inflows, market volatility as measured by the Market Volatility index (VIX), and the Federal Funds rate, i.e. the US monetary policy instrument. Data on gross capital flows are quarterly, and are taken from the IMF Balance of Payments Statistics. VIX data are at monthly frequency as well as data on the Federal Funds rate.[10]

As in the recent literature on capital flow dynamics, we focus on gross inflows, as opposed to net flows, because the size and variability of the former are much larger than those of net flows as shown by the two aggregates for the US economy reported in Figure 3 (together with the US outflows expressed as a percentage of GDP).[11] Second, changes in gross capital inflows characterize periods of crisis, such as the global financial crisis and the dot-com bubble at the beginning of the new millennium.

We measure financial market volatility by the VIX of the Chicago Board Options Exchange. Formally, the VIX index represents the option-implied expected volatility on the S&P500 index with a horizon of 30 calendar days or, equivalently, 22 trading days. The VIX can be viewed as an indicator of global risk as it captures overall "economic uncertainty" or "risk", including both the riskiness of financial assets and investor risk aversion (see Forbes and Warnock, 2012; Passari and Rey, 2015). Forbes and Warnock (2012) show that the VIX is the most consistently significant variable in predicting extreme capital flow episodes. Higher levels of the VIX are positively correlated with stops and retrenchments and negatively correlated with surges and flights. Miranda-Agrippino and Rey (2015) find that risky asset prices (equities, corporate bonds) around the world are largely driven by one global factor that is tightly negatively related to the VIX.[12]

The last endogenous variable in our model is the Federal Funds rate that is the policy instrument that the Federal Reserve uses to conduct its monetary policy.[13] The monetary policy reaction function that determines the Federal Funds rate is often thought to have two components: ($i$) a systematic, anticipated, reaction to key macroeconomic variables (inflation, output gap, etc.) in the spirit of the Taylor (1993) rule, and; ($ii$) an unanticipated "monetary policy shock".

[10]Further details on the specification of the MIDAS-SVAR model are provided in Section 4.1.

[11]The terminology can be confusing. Gross inflows are the difference between foreign purchases of domestic assets less foreign sales of domestic assets.

[12]The global factor explains one-fourth of the variance of the returns of the 858 risky asset in the sample.

The first step is to identify the monetary policy shocks, i.e. the structural shocks to the Federal Funds rate, that will be used to investigate the transmission of monetary policy to gross capital flows and the VIX through impulse response functions. The assumption generally made for identification in a SVAR framework is that the Federal Funds rate does not contemporaneously affect inflation and the output gap (as well as other macroeconomic aggregates).[14] This implies that news to the inflation rate and the output gap are exogenous to the Federal Funds rate and monetary policy shocks can be derived from a regression of the Federal Funds rate on the contemporaneous and lagged values of such variables. We exploit this standard identification strategy to estimate monetary policy shocks by including in the equation for the Federal Funds rate, as exogenous variables, the contemporaneous value and two monthly lags of the inflation rate and the growth rate of industrial production, the latter as a proxy for the output gap.[15] In the Technical Supplement, we show that our findings are robust to the inclusion of such variables as endogenous variables in the VAR.

Finally, we include in all equations, as exogenous control variables, lagged EU and Japanese short-term interest rates. In fact, there is substantial evidence showing that European interest rates are affected by US monetary policy (see, among others, Favero and Giavazzi (2008)) and the same is probably true for the interest rates of many other countries (see Passari and Rey, 2015). However, to the extent that foreign interest rates react to US policy shocks, we expect US capital inflows to slow down as other low-risk countries follow the same monetary policy.

As many macroeconomic variables are available only quarterly, SVAR models used to analyze the transmission of monetary policy shocks to the real economy are often based on this sampling frequency. A preliminary analysis of our model based on quarterly data shows a positive, but statistically not significant, reaction of capital inflows to a positive shock of the Federal Funds rate, as well as a feeble negative response of capital inflows to a VIX shock.[16] In the next Section we compare these results to those obtained from the MIDAS-SVAR estimation of the same model in order to asses the presence of distortions due to the "temporal aggregation bias"

[13]More precisely, US monetary policy is conducted by setting an operating target for the Federal Funds rate; i.e. the overnight interest rate on the interbank market for excess reserves (or Federal Funds).

[14]The literature on the effects of monetary policy shocks to the real economy is huge and mainly differentiates according to the empirical strategy used for the identification of the macroeconomic shocks. See among many others Christiano, Eichenbaum, and Evans (2005) for recursive identification schemes, while Bacchiocchi and Fanelli (2015) and Bacchiocchi, Castelnuovo, and Fanelli (2017) for non-recursive identification schemes and references therein for alternative approaches.

[15]The inflation rate and the growth rate of industrial production are also included as exogenous variables in the other two equations of the VAR. Being inflation and the growth rate of industrial production observed monthly, in each equation they enter contemporaneously and up to two monthly lags.

[16]The details of a preliminary analysis with the quarterly VAR are described in the Technical Supplement.

(see e.g. Christiano and Eichenbaum, 1987; Bayar, 2014; Foroni and Marcellino, 2014). In fact, as monthly realizations of the interest rate and the VIX may contain useful information to unveil the structural and dynamic relationships among the variables, this comparison allows us to verify the usefulness of MIDAS-VAR estimation.

### 4.1 The MIDAS-VAR reduced-form model

Consider the MIDAS-VAR model

$$A(L)\,\tilde{x}(t) = C(L)\,\tilde{z}(t) + \tilde{u}(t) \tag{18}$$

with $\tilde{x}(t) = \left(x_H(t,1)', x_H(t,2)', x_H(t,3)', x_L(t)'\right)'$, and more precisely

$$\begin{aligned} x_H(t,j) &= \begin{pmatrix} i(t,j) \\ vix(t,j) \end{pmatrix} \qquad j = 1,..,3 \\ x_L(t) &= k(t) \end{aligned} \tag{19}$$

where $i(t,j)$ and $vix(t,j)$ are the Fed Funds rate and the VIX, respectively, all observed at the $j$-th month of quarter $t$, while $k(t)$ measures the gross capital inflows-GDP ratio for the quarter $t$. A set of exogenous variables $\tilde{z}_t$, as discussed above, is included to identify the structural shocks. In particular, we consider the inflation rate $\pi(t,j)$ and the growth rate of industrial production $\Delta ip(t,j)$, with $j = 1,..,3$, observed at monthly frequency.[17] The inclusion of these variables allows us to identify the monetary policy shocks.

The reduced-form MIDAS-VAR can thus be modeled as in Eq. (1) where $\Sigma_{\tilde{u}}$ is the covariance matrix of the residuals, as in Eq. (3). The time index remains the quarter $t$ and the reduced form can be treated as a traditional VAR model in which the high frequency variables enter at all monthly frequencies. This, as shown below, helps to identify the different structural shocks hitting the dependent variables in each month within the quarter.

The optimal number of lags (in quarters) can be obtained through the standard approach. The Akaike and Bayesian information criteria, joint with the standard Lagrange Multiplier tests for the autocorrelation and the multivariate normality of the residuals suggest to include simply two lags.

### 4.2 The MIDAS-SVAR response of capital inflows to monetary policy and market volatility

The AB-MIDAS-SVAR model provides a very general approach to investigate the transmission of policy and non-policy shocks in a mixed-frequency data framework. As discussed in Section

[17] In this specification the two exogenous variables enter without lags within the same quarter $t$, i.e. without considering lags $t-1$, $t-2$, .... Including such lags, however, does not change the empirical results presented in this section.

2.1, the covariance matrix of the residuals $\Sigma_{\tilde{u}}$ hides all contemporaneous relations among the high- and low-frequency variables, the within quarter relations between low- and high-frequency variables and the within quarter dynamics between $x_H(t,i)$ and $x_H(t,j)$, with $i > j$. These relations will be made explicit through the $A$ matrix. The other contemporaneous relations, instead, are specified in the $B$ matrix that shows the simultaneous effect of the structural shocks among the variables, and within the quarter. The exactly identified structural form becomes:

$$\begin{pmatrix} 1 & & & & & & \\ & 1 & & & & & \\ * & * & 1 & & & & \\ * & * & & 1 & & & \\ * & * & * & * & 1 & & \\ * & * & * & * & & 1 & \\ & & & & & & 1 \end{pmatrix} \begin{pmatrix} u^{i}(t,1) \\ u^{vix}(t,1) \\ u^{i}(t,2) \\ u^{vix}(t,2) \\ u^{i}(t,3) \\ u^{vix}(t,3) \\ u^{k}(t) \end{pmatrix} = \begin{pmatrix} * & & & & & & \\ * & * & & & & & \\ & & * & & & & \\ & & * & * & & & \\ & & & & * & & \\ & & & & * & * & \\ * & * & * & * & * & * & * \end{pmatrix} \begin{pmatrix} \varepsilon^{mp}(t,1) \\ \varepsilon^{v}(t,1) \\ \varepsilon^{mp}(t,2) \\ \varepsilon^{v}(t,2) \\ \varepsilon^{mp}(t,3) \\ \varepsilon^{v}(t,3) \\ \varepsilon^{k}(t) \end{pmatrix} \quad (20)$$

where asterisks ($*$) denote unrestricted coefficients and empty entries correspond to zeros. The previous relation in Eq. (20), using the estimated residuals from the MIDAS-VAR in Eq. (18), allows us to identify the structural shocks $\varepsilon^{mp}(t,j)$, $\varepsilon^{v}(t,j)$ and $\varepsilon^{k}(t)$ that represent the Federal Funds rate shock, the market volatility, VIX, shock and the capital inflow shock, respectively. In particular, a recursive structure is assumed, in which VIX shocks affect the policy rate after one month, while shocks to the Federal Funds rate contemporaneously impact on the VIX. This set of assumptions, according to which the financial market volatility indicator, in a recursive scheme, is ordered after the policy rate, is in line with Bekaert, Hoerova, and Lo Duca (2013), among many others. Interestingly, the mixed frequency nature of the variables allows us to identify the high frequency structural shocks hitting the low frequency variables $m$ times ($m = 3$ in our empirical analysis) within the same quarter $t$. This is a main contribution of our methodology.

The "relatively reduced" dimensionality of the model makes the ML estimator, generally used in the traditional SVAR literature, easily implementable and allows hypothesis testing on the restrictions in the $A$ and $B$ matrices in Eq. (20), as well as those on the dynamics of the model, to behave as standard LR tests.[18]

In Section 3 we proposed a test for investigating whether the MIDAS-SVAR model is effectively more powerful than a traditional SVAR model, both for the dynamics part and the structural part of the model. Starting from the reduced form, the log-likelihood test statistic is $LR = -2(-132.114 - 37.881) = 339.991$ and, when compared with its asymptotic distribution under the null, i.e. a $\chi^2_{(36)}$, it leads to a clear rejection of the null hypothesis with a p-value practically equal to 0. Therefore, the test strongly suggests that the MIDAS-SVAR model provides much more accurate results than the traditional SVAR using low-frequency variables only. The details concerning the test implementation are provided in Appendix C.

[18]The estimates become quasi-ML when the assumption of a Gaussian likelihood is not supported by the data. As a consequence, all LR tests should be interpreted as quasi-LR tests.

Moreover, if we focus on the structural form, we can perform different tests in order to check whether the information contained in the monthly variables really matters in terms of a) the identification of the shocks and b) the propagation of such shocks to the dynamics of US capital inflows. Concerning the former, we first check whether the interest rate and the VIX at time $(t,1)$ help to identify the monetary policy shock at time $(t,2)$ and, jointly, interest rate and VIX at time $(t,1)$ and $(t,2)$ help identifying the monetary policy shock at time $(t,3)$. This assumption corresponds to the following structure:

$$\begin{pmatrix} 1 & & & & & & \\ & 1 & & & & & \\ \mathbf{0} & \mathbf{0} & 1 & & & & \\ * & * & & 1 & & & \\ \mathbf{0} & \mathbf{0} & \mathbf{0} & \mathbf{0} & 1 & & \\ * & * & * & * & & 1 & \\ & & & & & & 1 \end{pmatrix} \begin{pmatrix} u^{i}(t,1) \\ u^{vix}(t,1) \\ u^{i}(t,2) \\ u^{vix}(t,2) \\ u^{i}(t,3) \\ u^{vix}(t,3) \\ u^{k}(t) \end{pmatrix} = \begin{pmatrix} * & & & & & & \\ * & * & & & & & \\ & & * & & & & \\ & & * & * & & & \\ & & & & * & & \\ & & & & * & * & \\ * & * & * & * & * & * & * \end{pmatrix} \begin{pmatrix} \varepsilon^{mp}(t,1) \\ \varepsilon^{v}(t,1) \\ \varepsilon^{mp}(t,2) \\ \varepsilon^{v}(t,2) \\ \varepsilon^{mp}(t,3) \\ \varepsilon^{v}(t,3) \\ \varepsilon^{k}(t) \end{pmatrix}.$$

However, the likelihood test statistic, $LR = 372.452$, suggests to strongly reject the six overidentifying restrictions with a p-value practically equal to 0.

Similarly, we can check whether, jointly, the monetary policy shock at time $(t,1)$ helps to identify the volatility shock at time $(t,1)$, the endogenous variables and the structural shocks at time $(t,1)$ helps identifying the volatility shock at time $(t,2)$ and whether the endogenous variables and the structural shocks at time $(t,1)$ and $(t,2)$ help to identify the volatility shock at time $(t,3)$, i.e.

$$\begin{pmatrix} 1 & & & & & & \\ & 1 & & & & & \\ * & * & 1 & & & & \\ \mathbf{0} & \mathbf{0} & & 1 & & & \\ * & * & * & * & 1 & & \\ \mathbf{0} & \mathbf{0} & \mathbf{0} & \mathbf{0} & & 1 & \\ & & & & & & 1 \end{pmatrix} \begin{pmatrix} u^{i}(t,1) \\ u^{vix}(t,1) \\ u^{i}(t,2) \\ u^{vix}(t,2) \\ u^{i}(t,3) \\ u^{vix}(t,3) \\ u^{k}(t) \end{pmatrix} = \begin{pmatrix} * & & & & & & \\ \mathbf{0} & * & & & & & \\ & & * & & & & \\ & & \mathbf{0} & * & & & \\ & & & & * & & \\ & & & & \mathbf{0} & * & \\ * & * & * & * & * & * & * \end{pmatrix} \begin{pmatrix} \varepsilon^{mp}(t,1) \\ \varepsilon^{v}(t,1) \\ \varepsilon^{mp}(t,2) \\ \varepsilon^{v}(t,2) \\ \varepsilon^{mp}(t,3) \\ \varepsilon^{v}(t,3) \\ \varepsilon^{k}(t) \end{pmatrix}$$

As for the previous case, this null hypothesis (nine overidentifying restrictions) is strongly rejected by the data ($LR = 209.978$ and p-value=0.00).

Concerning the latter kind of tests, focusing on the transmission of the structural shocks, we first check the null hypothesis that only the monetary policy and volatility shocks occurring in the first month of the quarter contribute to explain the dynamics of US capital inflows. This hypothesis imposes four zero restrictions on the last row of the $B$ matrix in Eq. (20). The related likelihood ratio test statistic, $LR = 11.287$, suggests to reject (at 10% and 5% significant levels) the null hypothesis with a p-value= 0.02. Other two interesting null hypotheses are that the impact of interest rate shocks first, and VIX shocks then, have the same impact across the different months within the quarter on the dynamics of capital inflows. Both hypotheses impose two equality restrictions in the last row of the $B$ matrix in Eq. (20). The conclusions are opposite. While monetary policy shocks have statistically different effects on US capital inflows within the quarter ($LR = 13.915$, with a p-value=0.00), the on impact response of capital inflows to volatility shocks seems to be statistically equivalent over the three months of the quarter ($LR = 1.178$, with a p-value=0.56).

### 4.3 Impulse responses and variance decomposition

The impulse response functions (IRFs) of capital inflows, $k(t)$, to the different types of shocks are shown in Figure 5. As discussed above, the low frequency variable $k(t)$ is expected to respond to interest-rate and VIX shocks occurring in all three months within the quarter. Impulse responses are displayed in the first row of Figure 5 for shocks to the Federal Funds rate and in the second row for shocks to the VIX. The bottom panel reports the response of capital inflows to a shock to themselves.

The first interesting result is that the 'on-impact' effect on capital inflows of an unanticipated increase in the Fed Funds rate is different depending on the month it happens within the quarter, though the dynamic response is similar in the three cases. The effect is positive and significant when the interest-rate shock occurs in the first month of the quarter, positive but significant only after two quarters when the shock occurs in the second month, while it becomes negative when the shock takes place in the third month, i.e. at the end of the quarter. In other words, an unexpected monetary contraction, i.e. a positive shock the Fed Funds rate, has a positive effect on capital inflows when it takes place at the beginning of the quarter while the effect is negative at the end of the quarter. Thus, it appears that monetary policy shocks take time to display their effect on capital flows. To the extent that changes in interest rates are persistent, a shock that occurs at the beginning of the quarter, and lasts over three months, is expected to have a larger effect on capital inflows as it affects the net sale of assets over the entire three-month period over which they are measured. On the other hand, an interest rate shock at the end of the quarter barely affects capital inflows within that quarter since the latter are mostly determined by the market conditions prevailing in the previous two months. The delayed effect of interest rate shocks occurring in the second month of the quarter is consistent with this interpretation. While asset prices and exchange rates immediately react to interest rate shocks, US capital inflows, i.e. the net sale of US assets to foreign residents, being a flow variable, increase with the sampling period. An increase in the interest rate occurring at the beginning of the quarter –and lasting until the end of the quarter– that boosts sales of US assets can have a sizable effect on capital flows simply because such sales cumulate over the entire three-month period.

This interpretation requires interest rate shocks to be persistent; if they were short lived, in particular if they just lasted one month, their impact on capital flows would be the same independently of the month they occurred. Figure 6 shows the response of the Fed Funds rate to its own shock. A monetary policy shock has a rather persistent effect on the interest rate. A shock that occurs at the beginning of the quarter produces a significant effect on the following two months too, before capital inflows data are collected. Actually, the response of the Fed Funds rate is hump-shaped reaching a peak after two months which suggests the possibility of a delayed reaction of capital flows. The result that interest rate shocks have different effects depending on their timing within the quarter explains the evidence reported in the Technical Supplement for the quarterly VAR that monetary policy has no (or at most very weak) impact on capital inflows. In fact, aggregating the three impulse responses of Figure 5 (left panel) makes the overall quarterly effect quite weak.

The middle row of Figure 5 shows the different responses of capital inflows to VIX shocks for each of the months in the quarter. The results are qualitatively similar to those obtained with the quarterly SVAR described in the Technical Supplement but the negative impact of VIX shocks on capital inflows is stronger. In particular, financial market volatility seems to have an immediate and significant effect on capital inflows, independently of the month the shock takes place, though the strongest impact is observed when the shock occurs in the second month.

Figures 6-8 complete the first set of results. In Figure 6 we show the response of the Federal Funds rate to a shock on itself and to shocks to previous months rates. Figure 7 reports the responses of the VIX to interest rate shocks. Interestingly, a shock to the Fed Funds rate occurring in the first month strongly confirms the results in Bekaert, Hoerova, and Lo Duca (2013). In fact, although we do not decompose the VIX into a risk aversion indicator and an uncertainty proxy, we find that a monetary contraction at the beginning of the quarter immediately reduces the volatility observed on financial markets, while the effect turns positive and strongly significant three-four quarters after the shock. A similar but less significant pattern is observed when the interest rate shock occurs in the second and third month within the quarter.

Figure 8 reports the reaction of the Fed Funds rate to VIX shocks. In line with Bekaert, Hoerova, and Lo Duca (2013), an increase in financial volatility (or in risk aversion, in their setting) prompts a monetary easing, especially when the VIX shock occurs in the second month of the quarter. However, such effect is not statistically significant.

In Table 1 we report the Forecast Error Variance Decomposition (FEVD) for the capital inflows subject to monetary policy and volatility shocks in the AB-MIDAS-SVAR model (panel b), calculated at different horizons (0, 1, 4, 8, 20 quarters). Such results are compared to those obtained through the aggregate quarterly SVAR discussed in the Technical Supplement (panel a). In the AB-MIDAS-SVAR model, shocks to the Federal Funds rate account for about the double of the variance of capital flows compared with the quarterly SVAR model after 20 quarters, while this ratio is up to ten times for the response on impact. Evidence in favor of the MIDAS-SVAR with respect to the traditional SVAR is substantially confirmed when comparing the effects of VIX shocks to those in the quarterly SVAR.

All the previous results show that the MIDAS-SVAR model performs significantly better than the quarterly SVAR: All the identified structural shocks explain a much larger part of the forecast error variance and provide richer impulse responses by exploiting the dynamics within the quarter.

# 5 Robustness checks

In this section we provide a rich set of robustness checks regarding the identification of the shocks, the impact of other control variables and the possible effects of the zero lower bound. All the details of these further investigations, as well as the IRFs and the FEVDs are reported in the Technical Supplement.

## 5.1 Robustness: different identification scheme

In the specification of the MIDAS-SVAR in Section 4.2, a recursive structure is assumed in which the VIX shocks affect the interest rate after one month, while the interest rate shocks contemporaneously impact on the VIX. This set of assumptions, though in line with Jurado, Ludvigson, and Ng (2015), is at odds with Bloom (2009) and Caggiano, Castelnuovo, and Groshenny (2014) who arrange the VIX as first in the SVAR specification.[19]

In order to validate our findings we repeat the empirical analysis inverting the ordering of the monthly variables in the MIDAS-SVAR. The resulting IRFs, reported in Figure TS.7 in the Technical Supplement, substantially confirm the main findings of Section 4.3.

## 5.2 Robustness: Endogenous economic activity indicator and inflation rate

In the empirical analysis, in order to control for parameters proliferation, the growth rate of industrial production and the inflation rate are considered as exogenous variables. This choice, though consistent with the identification of the monetary policy shock, is questionable in that such variables are usually included as endogenous in VAR specifications, so as to determine the growth rate of industrial production and inflation jointly with the interest rate.

The analysis presented in Section 4 is replicated with the inclusion of these monthly variables as endogenous. In particular, the identification structure is the same as in Eq. (20), with the growth rate of industrial production and inflation, $\Delta ip(t,j)$ and $\pi(t,j)$, ordered just before the interest rate $i(t, j)$, with $j = 1, \ldots, 3$. The results, reported in the Technical Supplement, show no substantial differences in the responses of capital inflows to interest-rate and VIX shocks, for each month within the quarter.

[19]Ludvigson, Ma, and Ng (2017), Angelini, Bacchiocchi, Caggiano, and Fanelli (2017) and Carriero, Clark, and Marcellino (2017), in three recent contributions, question the exogenous role or the endogenous response of financial uncertainty indicators, like the VIX, towards the real economy. Prevailing the former or the latter of these two causality directions might have important implications on the short-run restrictions imposed to identify the structural latent shocks.

### 5.3 Robustness: US capital inflows at the 'zero lower bound'

Just after the global financial crisis, the Federal Reserve started to implement unconventional monetary policies rather than acting on the Federal Funds rate that remained practically constant around zero. The absence of variability in the Federal Funds rate, however, is not accompanied by a similar path of gross capital inflows. In order to check whether the alternative policy instruments change the transmission of monetary policy to capital flows we repeat the empirical analysis by substituting the Federal Funds rate with the shadow rate proposed by Wu and Xia (2016). This new variable, shown in Figure TS.8 in the Technical Supplement, perfectly mimics the dynamics of the Federal Funds rate up to 2009 and, then, becomes negative to account for the further monetary expansion that was carried out with unconventional instruments. The estimated impulse responses for capital inflows, reported in the Technical Supplement, are very similar to the main findings presented in Section 4.3.

### 5.4 Robustness: controlling for further exogenous variables

In Section 4 we discussed the importance of considering economic activity and inflation to identify monetary policy shocks. However, other variables may be relevant to explain the dynamics of the VIX and US capital inflows. Thus, we include, as exogenous variables: an indicator of the global business cycle measured by the growth rate of aggregate industrial production in OECD+BRICST countries (Brazil, Russia, India, China, South Africa and Turkey); an indicator of the US stock market, as measured by the return on the S&P500 price index[20]; the 10-year interest rate on US Treasuries; the effective exchange rate, and; an alternative indicator for the US business cycle measured by the monthly civilian unemployment rate. To avoid endogeneity issues we rely on the lagged values of these monthly variables. All the results are practically identical to those presented in Section 4.3, and are available from the authors upon request.

## 6 Concluding remarks

We presented a new framework for performing structural analysis with mixed-frequency data. This model, that we called MIDAS-SVAR, fully exploits the different frequencies of the variables, and its specification appears as a generalization of standard VARs. We have also proposed a test of hypothesis for statistically evaluating the gains of using mixed frequencies with respect to low frequency standard VARs. The asymptotic distribution of the test statistic is standard, while in small samples the performance of the test has been evaluated through a set of Monte Carlo experiments. For small- and medium-scale MIDAS-VARs the gains are enormous when the DGP is based on real data. Moreover, the gains reveal to be consistent even when the DGP is quite close to a standard low frequency VAR.

[20]Qualitatively identical results are obtained relying on the MSCI-World Index or running the model using the levels instead of the growth rates of variables.

Using this econometric framework we presented new evidence on the effects of monetary policy and financial market volatility on US capital inflows. In particular, we showed that the (so far) weak evidence of an effect of interest rate shocks on US capital inflows is due to the different impact that such shocks have depending on the month in which they occur within the quarter. The MIDAS-VAR allows to highlight these different effects. Specifically, an interest rate shock has a strong positive impact on capital inflows when it occurs in the first two months of the quarter, while the effect is negative when the shock hits the economy in the last month of the quarter.

# A Appendix: Mapping from MIDAS-VAR to VAR: Some details and further results

Consider the MIDAS-VAR models in Eq. (9) with potential exogenous variables included:

$$\begin{pmatrix} x_H^1(t,1) \\ x_H^1(t,2) \\ x_H^1(t,3) \\ x_L^2(t) \end{pmatrix} = \begin{pmatrix} A_{11}^1 & A_{12}^1 & A_{13}^1 & A_1^1 \\ A_{21}^1 & A_{22}^1 & A_{23}^1 & A_2^1 \\ A_{31}^1 & A_{32}^1 & A_{33}^1 & A_3^1 \\ A_{L1}^1 & A_{L2}^1 & A_{L3}^1 & A_L^1 \end{pmatrix} \begin{pmatrix} x_H^1(t-1,1) \\ x_H^1(t-1,2) \\ x_H^1(t-1,3) \\ x_L^2(t-1) \end{pmatrix} + \\ + \begin{pmatrix} C_{11}^1 & C_{12}^1 & C_{13}^1 & C_1^1 \\ C_{21}^1 & C_{22}^1 & C_{23}^1 & C_2^1 \\ C_{31}^1 & C_{32}^1 & C_{33}^1 & C_3^1 \\ C_{L1}^1 & C_{L2}^1 & C_{L3}^1 & C_L^1 \end{pmatrix} \begin{pmatrix} z_H(t,1) \\ z_H(t,2) \\ z_H(t,3) \\ z_L(t) \end{pmatrix} + \begin{pmatrix} u_H^1(t,1) \\ u_H^1(t,2) \\ u_H^1(t,3) \\ u_L^2(t) \end{pmatrix} \tag{21}$$

where $z_H(t,i)$ is the vector of high-frequency exogenous variables for the $i$-th month of quarter $t$ and $z_L(t)$ is the vector of low-frequency exogenous variables at quarter $t$. If the aggregation scheme consists in taking the first observation of the quarter only, the $G$ matrix is the one reported in Eq. (11) and the associated null hypothesis is as described in Eqs. (12)-(15) for testing the equivalence between the two specifications (without exogenous variables, for simplicity). If the model specification allows for exogenous regressors, the previous null hypothesis must be completed with

$$\begin{aligned} H_0^1 &: A_{12}^1 = A_{13}^1 = 0 \\ H_0^2 &: A_{L2}^1 = A_{L3}^1 = 0. \end{aligned} \tag{22}$$

A very similar situation occurs when the aggregation scheme for high-frequency variables reduces to select the last observation of the quarter.

## A.1 Proofs

*Proof of Proposition 2.* Let $\ddot{x}(t) = G\tilde{x}(t)$, with

$$A(L)\tilde{x}(t) = \tilde{u}(t),$$

premultiplying by $G$

$$\begin{aligned} GA(L)\tilde{x}(t) &= G\tilde{u}(t) \\ G(I_{\tilde{n}} - A_1L - \ldots - GA_pL^p)\tilde{x}(t) &= G\tilde{u}(t) \\ G\tilde{x}(t) - GA_1\tilde{x}(t-1) - \ldots - GA_p\tilde{x}(t-p) &= G\tilde{u}(t) \end{aligned}$$

with $GA_i\tilde{x}(t-i) = \ddot{A}_i\ddot{x}(t-i), i = 1, 2, \ldots, p$, then

$$E[G\tilde{x}(t)|\tilde{x}(t-1), \ldots, \tilde{x}(t-p)] = \ddot{A}_1\ddot{x}(t-1) + \ldots + \ddot{A}_p\ddot{x}(t-p).$$

□

*Proof of Proposition 3.* Given the SVAR for $\ddot{x}(t) = G\tilde{x}(t)$

$$A\ddot{u}(t) = B\ddot{\varepsilon}(t),$$

premultiplying the MIDAS-SVAR model by $G$

$$GA\tilde{u}(t) = GB\tilde{\varepsilon}(t)$$

with $GA\tilde{u}(t) = A\ddot{u}(t)$ and $GB\tilde{\varepsilon}(t) = B\ddot{\varepsilon}(t)$ the two models are equivalent. □

## A.2 Different aggregation schemes

The use of the selection matrix $G$ allows to evaluate the differences between MIDAS-SVARs and traditional SVARs when alternative aggregations of the high frequency variables occur. Consider, for example, the quarterly data obtained by cumulating monthly data:

$$\ddot{x}_L^1(t) = \sum_{i=1}^{3} x_H^1(t,i), \tag{23}$$

then, define the selection $G$ matrix as follows

$$G = \begin{pmatrix} I_{n_H} & I_{n_H} & I_{n_H} & 0 \\ 0 & 0 & 0 & I_{n_L} \end{pmatrix} \tag{24}$$

and premultiply both sides of Eq. (9) by $G$. After some algebra, we obtain that

$$\begin{pmatrix} \ddot{x}_L^1(t) \\ \ddot{x}_L^2(t) \end{pmatrix} = \begin{pmatrix} \ddot{A}_{11}^1 & \ddot{A}_{12}^1 \\ \ddot{A}_{21}^1 & \ddot{A}_{22}^1 \end{pmatrix} \begin{pmatrix} x_H^1(t-1,1) \\ x_H^1(t-1,2) \\ x_H^1(t-1,3) \\ x_L^2(t-1) \end{pmatrix} + \begin{pmatrix} \ddot{C}_{11}^1 & \ddot{C}_{12}^1 \\ \ddot{C}_{21}^1 & \ddot{C}_{22}^1 \end{pmatrix} \begin{pmatrix} z_H^1(t-1,1) \\ z_H^1(t-1,2) \\ z_H^1(t-1,3) \\ z_L^2(t) \end{pmatrix} + \begin{pmatrix} \ddot{u}_L^1(t) \\ \ddot{u}_L^2(t) \end{pmatrix} \tag{25}$$

with

$$\begin{array}{lcl}\ddot{A}^1_{11} & = & \left( A^1_{11}+A^1_{21}+A^1_{31} \quad A^1_{12}+A^1_{22}+A^1_{32} \quad A^1_{13}+A^1_{23}+A^1_{33} \right) \\ \ddot{A}^1_{21} & = & \left( A^1_{L1} \quad A^1_{L2} \quad A^1_{L3} \right) \\ \ddot{A}^1_{12} & = & A^1_1+A^1_2+A^1_3 \\ \ddot{A}^1_{22} & = & A^1_L \\ \ddot{C}^1_{11} & = & \left( C^1_{11}+C^1_{21}+C^1_{31} \quad C^1_{12}+C^1_{22}+C^1_{32} \quad C^1_{13}+C^1_{23}+C^1_{33} \right) \\ \ddot{C}^1_{21} & = & \left( C^1_{L1} \quad C^1_{L2} \quad C^1_{L3} \right) \\ \ddot{C}^1_{12} & = & C^1_1+C^1_2+C^1_3 \\ \ddot{C}^1_{22} & = & C^1_L \end{array} \tag{26}$$

When there are no exogenous variables, the MIDAS-VAR and the VAR models described in Eq. (9) and Eq. (10), respectively, are equivalent if

$$\begin{array}{lcl} H^1_0 & : & \left(A^1_{11}+A^1_{21}+A^1_{31}\right)=\left(A^1_{12}+A^1_{22}+A^1_{32}\right)=\left(A^1_{13}+A^1_{23}+A^1_{33}\right) \\ H^2_0 & : & A^1_{L1}=A^1_{L2}=A^1_{L3}. \end{array} \tag{27}$$

In the case exogenous variables are included, as in the empirical application presented in the paper, the null hypothesis for testing the mapping between the two specifications must include the following relations

$$\begin{array}{lcl} H^3_0 & : & \left(C^1_{11}+C^1_{21}+C^1_{31}\right)=\left(C^1_{21}+C^1_{22}+C^1_{32}\right)=\left(C^1_{31}+C^1_{32}+C^1_{33}\right) \\ H^4_0 & : & C^1_{L1}=C^1_{L2}=C^1_{L3}. \end{array} \tag{28}$$

Concerning the matching between SVAR and MIDAS-SVAR, the representations of the latter can be easily obtained by premultiplying both sides of Eq.(13) by the $G$ matrix defined before, regardless the presence or not of exogenous variables. The null hypothesis

$$\begin{array}{lcl} H^1_0 & : & (A_{11}+A_{21}+A_{31})=(A_{12}+A_{22}+A_{32})=(A_{13}+A_{23}+A_{33}) \\ H^2_0 & : & A_{L1}=A_{L2}=A_{L3} \\ H^3_0 & : & (B_{11}+B_{21}+B_{31})=(B_{12}+B_{22}+B_{32})=(B_{13}+B_{23}+B_{33}) \\ H^4_0 & : & B_{L1}=B_{L2}=B_{L3} \end{array} \tag{29}$$

immediately follows.

Extremely interesting, as it would reasonably be in many empirical applications, is the case where the high-frequency variables are aggregated differently with respect to their nature. As an example, the interest rate and the VIX could be selected at the beginning of the quarter, while inflation and the growth rate of industrial production could be aggregated by taking the quarter average. Were this the case, for an hypothetical vector of high-frequency variables defined by $x_H(t,j) = \left(i\,(t,j)\,, vix\,(t,j)\,, \pi\,(t,j)\,, \Delta ip\,(t,j)\right)'$, and one single low-frequency variable $k(t)$, the associated selection matrix would be

$$G = \left(\begin{array}{ccccc|ccccc|ccccc|c} 1 & 0 & 0 & 0 & 0 & 0 & 0 & 0 & 0 & 0 & 0 & 0 & 0 & 0 & 0 & 0 \\ 0 & 1 & 0 & 0 & 0 & 0 & 0 & 0 & 0 & 0 & 0 & 0 & 0 & 0 & 0 & 0 \\ 0 & 0 & 1 & 0 & 0 & 0 & 0 & 0 & 0 & 0 & 0 & 0 & 0 & 0 & 0 & 0 \\ 0 & 0 & 0 & 1/3 & 0 & 0 & 0 & 0 & 1/3 & 0 & 0 & 0 & 0 & 1/3 & 0 & 0 \\ 0 & 0 & 0 & 0 & 1/3 & 0 & 0 & 0 & 0 & 1/3 & 0 & 0 & 0 & 0 & 1/3 & 0 \\ \hline 0 & 0 & 0 & 0 & 0 & 0 & 0 & 0 & 0 & 0 & 0 & 0 & 0 & 0 & 0 & 1 \end{array}\right) \tag{30}$$

with related null hypothesis obtained through the same reasoning as explained before.

# B Appendix: Monte Carlo simulations, some details

In Section 3.3.2 we have presented some simulation experiments for a medium-scale MIDAS-VAR model with $n_H = 3$ high-frequency monthly variables and $n_L = 1$ quarterly variable with one single lag. In particular, **Model 1-**$H_1$ has been obtained through the following matrices of parameters:

$$\Sigma_{\tilde{u}} = \begin{pmatrix} 0.028 & -0.004 & -0.001 & 0.046 & -0.003 & 0.000 & 0.058 & -0.005 & -0.001 & 0.204 \\ -0.004 & 0.023 & 0.001 & -0.003 & 0.022 & 0.001 & 0.001 & 0.020 & 0.001 & -0.285 \\ -0.001 & 0.001 & 0.000 & -0.001 & 0.002 & 0.000 & 0.000 & 0.002 & 0.000 & 0.005 \\ 0.046 & -0.003 & -0.001 & 0.102 & -0.003 & -0.001 & 0.132 & -0.008 & -0.001 & 0.350 \\ -0.003 & 0.022 & 0.002 & -0.003 & 0.045 & 0.003 & 0.001 & 0.041 & 0.003 & -0.438 \\ 0.000 & 0.001 & 0.000 & -0.001 & 0.003 & 0.001 & 0.000 & 0.003 & 0.001 & -0.011 \\ 0.058 & 0.001 & 0.000 & 0.132 & 0.001 & 0.000 & 0.193 & -0.006 & -0.001 & 0.289 \\ -0.005 & 0.020 & 0.002 & -0.008 & 0.041 & 0.003 & -0.006 & 0.051 & 0.004 & -0.389 \\ -0.001 & 0.001 & 0.000 & -0.001 & 0.003 & 0.001 & -0.001 & 0.004 & 0.001 & -0.025 \\ 0.204 & -0.285 & 0.005 & 0.350 & -0.438 & -0.011 & 0.289 & -0.389 & -0.025 & 82.278 \end{pmatrix}. \tag{31}$$

and

$$A_1 = \begin{pmatrix} -0.008 & -0.041 & -1.350 & -0.392 & 0.093 & 0.176 & 1.379 & -0.105 & -0.181 & 0.002 \\ -0.149 & 0.060 & 0.845 & 0.219 & -0.210 & -2.055 & -0.053 & 1.116 & 0.832 & -0.002 \\ -0.007 & -0.002 & 0.121 & 0.012 & 0.018 & -0.156 & -0.005 & -0.003 & 0.836 & 0.000 \\ -0.134 & -0.208 & -3.217 & -0.519 & 0.295 & 0.174 & 1.599 & -0.293 & -0.578 & -0.002 \\ -0.146 & 0.194 & -0.101 & 0.170 & -0.179 & -1.323 & -0.019 & 0.877 & 0.683 & -0.001 \\ -0.023 & -0.014 & 0.136 & 0.036 & -0.002 & -0.036 & -0.014 & 0.027 & 0.582 & -0.001 \\ -0.241 & -0.310 & -3.159 & -0.544 & 0.422 & 1.560 & 1.726 & -0.356 & -3.078 & -0.003 \\ 0.110 & 0.283 & -0.411 & -0.122 & -0.292 & -0.408 & 0.011 & 0.811 & -0.300 & 0.001 \\ 0.007 & -0.015 & 0.113 & -0.002 & -0.020 & 0.248 & -0.007 & 0.037 & 0.275 & -0.001 \\ -1.200 & -3.441 & -73.371 & -1.461 & 12.762 & -66.768 & 1.423 & -8.138 & 4.622 & 0.376 \end{pmatrix} \tag{32}$$

Finally, the further models, **Model 2-$H_1$** and **Model 3-$H_1$**, are obtained through the same covariance matrix, $\Sigma_{\tilde{u}}$, and the following matrices of parameters

$$A_1 = \begin{pmatrix} 0.3 & 0 & 0 & 0.05 & 0.05 & 0.05 & -0.05 & -0.05 & -0.05 & 0 \\ 0 & 0.3 & 0 & 0.05 & 0.05 & 0.05 & -0.05 & -0.05 & -0.05 & 0 \\ 0 & 0 & 0.3 & 0.05 & 0.05 & 0.05 & -0.05 & -0.05 & -0.05 & 0 \\ 0 & 0 & 0 & 0.3 & 0 & 0 & 0 & 0 & 0 & 0 \\ 0 & 0 & 0 & 0 & 0.3 & 0 & 0 & 0 & 0 & 0 \\ 0 & 0 & 0 & 0 & 0 & 0.3 & 0 & 0 & 0 & 0 \\ 0 & 0 & 0 & 0 & 0 & 0 & 0.3 & 0 & 0 & 0 \\ 0 & 0 & 0 & 0 & 0 & 0 & 0 & 0.3 & 0 & 0 \\ 0 & 0 & 0 & 0 & 0 & 0 & 0 & 0 & 0.3 & 0 \\ 0.1 & 0.1 & 0.1 & 0.2 & 0.2 & 0.2 & 0.2 & 0.2 & 0.2 & 0.5 \end{pmatrix} \tag{33}$$

and

$$A_1 = \begin{pmatrix} 0.3 & 0 & 0 & 0.01 & 0.02 & 0.02 & -0.02 & -0.02 & -0.02 & 0 \\ 0 & 0.3 & 0 & 0.01 & 0.02 & 0.02 & -0.02 & -0.02 & -0.02 & 0 \\ 0 & 0 & 0.3 & 0.01 & 0.02 & 0.02 & -0.02 & -0.02 & -0.02 & 0 \\ 0 & 0 & 0 & 0.3 & 0 & 0 & 0 & 0 & 0 & 0 \\ 0 & 0 & 0 & 0 & 0.3 & 0 & 0 & 0 & 0 & 0 \\ 0 & 0 & 0 & 0 & 0 & 0.3 & 0 & 0 & 0 & 0 \\ 0 & 0 & 0 & 0 & 0 & 0 & 0.3 & 0 & 0 & 0 \\ 0 & 0 & 0 & 0 & 0 & 0 & 0 & 0.3 & 0 & 0 \\ 0 & 0 & 0 & 0 & 0 & 0 & 0 & 0 & 0.3 & 0 \\ 0.1 & 0.1 & 0.1 & 0.1 & 0.1 & 0.1 & 0.1 & 0.1 & 0.1 & 0.5 \end{pmatrix}, \tag{34}$$

respectively. The matrices of the parameters for generating **Model 1-$H_0$**, **Model 2-$H_0$** and **Model 3-$H_0$**, i.e. when the null hypothesis is true, are obtained starting from Eqs. (32)-(34) and imposing the restrictions presented in Eq. (12) and discussed in Section 3.3.2.

# C Appendix: MIDAS-SVAR *vs* SVAR, test implementation

This section is dedicated to the empirical implementation of the test. We consider that the low-frequency variables are obtained by adding high frequency variables. Alternative aggregation schemes are however discussed in Appendix A. We first focus on the reduced form model, and then discuss the structural form.

The reference null hypothesis is that reported in Eq. (27), and can be empirically implemented through a LR test. The reduced form is estimated both unrestrictedly and restrictedly and the two log-likelihood values are 37.881 and -132.114, respectively. The test statistic $LR = -2(-132.114 - 37.881) = 339.991$ asymptotically follows a $\chi^2_{(36)}$, with a related p-value practically equal to 0, strongly suggesting to reject the null hypothesis. Aggregating the high frequency series as in traditional VARs generates a loss of information that is statistically highly significant. The number of degrees of freedom is given by the number of restrictions on the parameters related to the dynamics of the VAR (first row of Eq.(27), i.e. 8 restrictions for each of the two lags), plus the restrictions on the relationships between low- and high-frequency variables (second row of Eq.(27), i.e. 4 restrictions for each of the two lags), plus a set of further 12 restrictions on the exogenous variables (the details are discussed in Appendix A).

Such incontrovertible result favoring the MIDAS-VAR model makes completely unnecessary the test on the matching between the structural part of the MIDAS-SVAR versus the SVAR model, whose implementation, however, would have followed the same LR principle, as postulated in Eq. (29).[21]

[21] As suggested in the Appendix A, one might be interested in alternative aggregation schemes. We have therefore tested whether the MIDAS-VAR model is comparable with the traditional VAR model obtained by taking the mean of the high-frequency variables instead of the first month of the quarter. The null hypothesis is strongly rejected. The results continue to confirm the better performance of the MIDAS-VAR model. See Appendix A for the details on the definition of the null hypothesis.

# D Appendix: Figures and Tables

Table 1: FEVD Quarterly SVAR and MIDAS-SVAR for capital flows

| Panel a. Quarterly SVAR | | | | | |
|---|---|---|---|---|---|
| $h$ | 0 | 1 | 4 | 8 | 20 |
| $i(t)$ | 1.25 | 2.66 | 3.17 | 4.31 | 12.75 |
| $vix(t)$ | 1.06 | 1.21 | 1.95 | 2.45 | 3.32 |
| $k(t)$ | 97.70 | 96.13 | 94.88 | 93.24 | 83.94 |
| Panel b. MIDAS-SVAR | | | | | |
| $h$ | 0 | 1 | 4 | 8 | 20 |
| $i(t,1)$ | 4.71 | 7.44 | 8.81 | 8.91 | 10.59 |
| $i(t,2)$ | 1.72 | 2.46 | 4.14 | 4.24 | 5.76 |
| $i(t,3)$ | 5.76 | 4.94 | 3.85 | 4.12 | 5.84 |
| $\sum i(t,m)$ | 12.19 | 14.84 | 16.80 | 17.27 | 22.19 |
| $vix(t,1)$ | 0.04 | 1.30 | 1.03 | 1.06 | 1.07 |
| $vix(t,2)$ | 1.88 | 2.21 | 3.60 | 3.54 | 3.48 |
| $vix(t,3)$ | 0.01 | 0.94 | 1.03 | 1.09 | 1.19 |
| $\sum vix(t,m)$ | 1.93 | 4.44 | 5.67 | 5.69 | 5.75 |
| $k(t)$ | 85.88 | 80.72 | 77.53 | 77.04 | 72.07 |

*Notes*: Forecast Error Variance Decomposition (FEVD) for the capital inflows subject to monetary policy shocks, economic and policy uncertainty shocks and capital shocks in the AB-MIDAS-SVAR model calculated at different horizons (0, 1, 4, 8, 20 quarters) (panel b), and FEVD obtained through the aggregate quarterly SVAR discussed in the Technical Supplement (panel a). The sample period is 1988:1-2013:3.

Figure 1: Monte Carlo simulations for small-scale MIDAS-VAR: pvalue plot (left panel) and size-power curve (right panel).

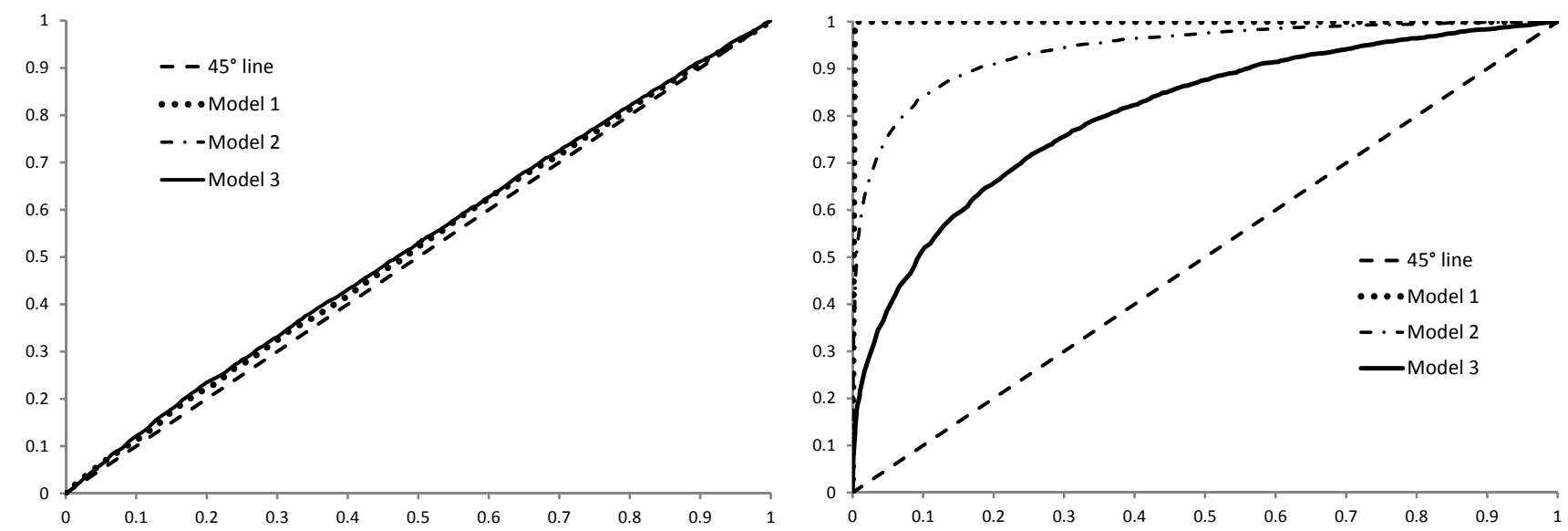

*Notes*: The DGPs small-scale MIDAS-VARs with $n_L = 1$ low-frequency variable and $n_H = 1$ high-frequency variable. The DGPs are presented in Section 3.3.1.

Figure 2: Monte Carlo simulations for medium-scale MIDAS-VAR: pvalue plot (left panel) and size-power curve (right panel).

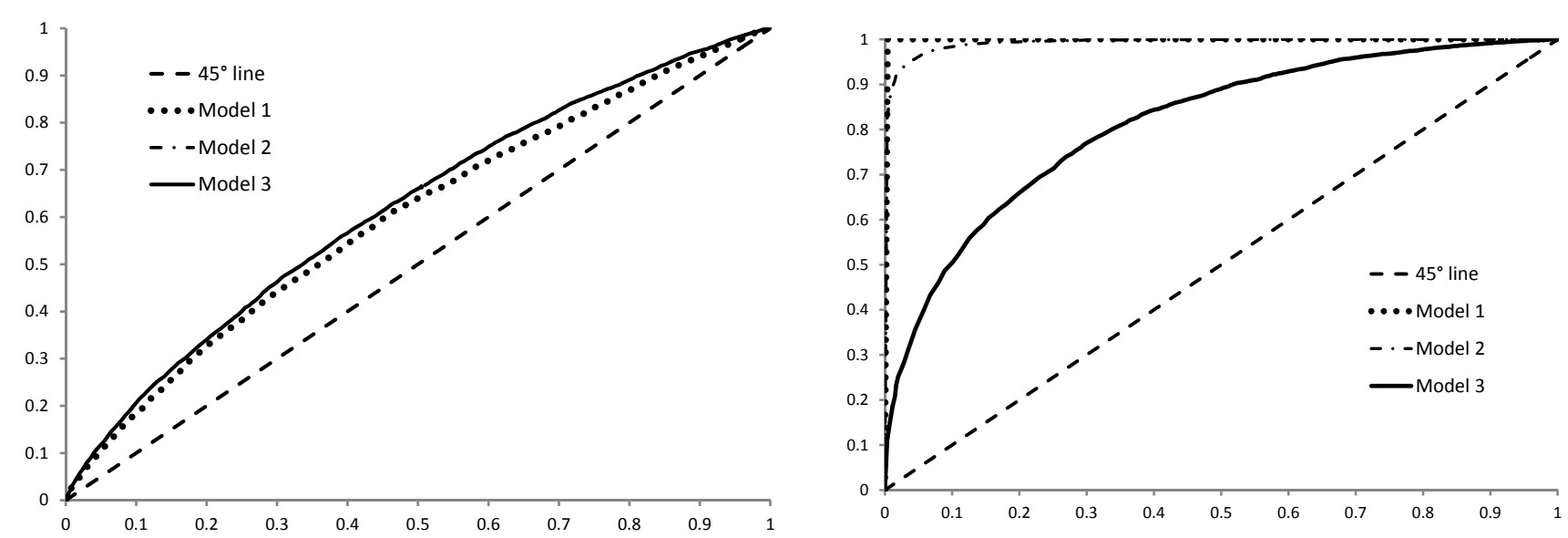

*Notes*: The DGPs medium-scale MIDAS-VARs with $n_L = 1$ low-frequency variable and $n_H = 3$ high-frequency variables. The DGPs are presented in Section 3.3.2.

Figure 3: US capital flows dynamics.

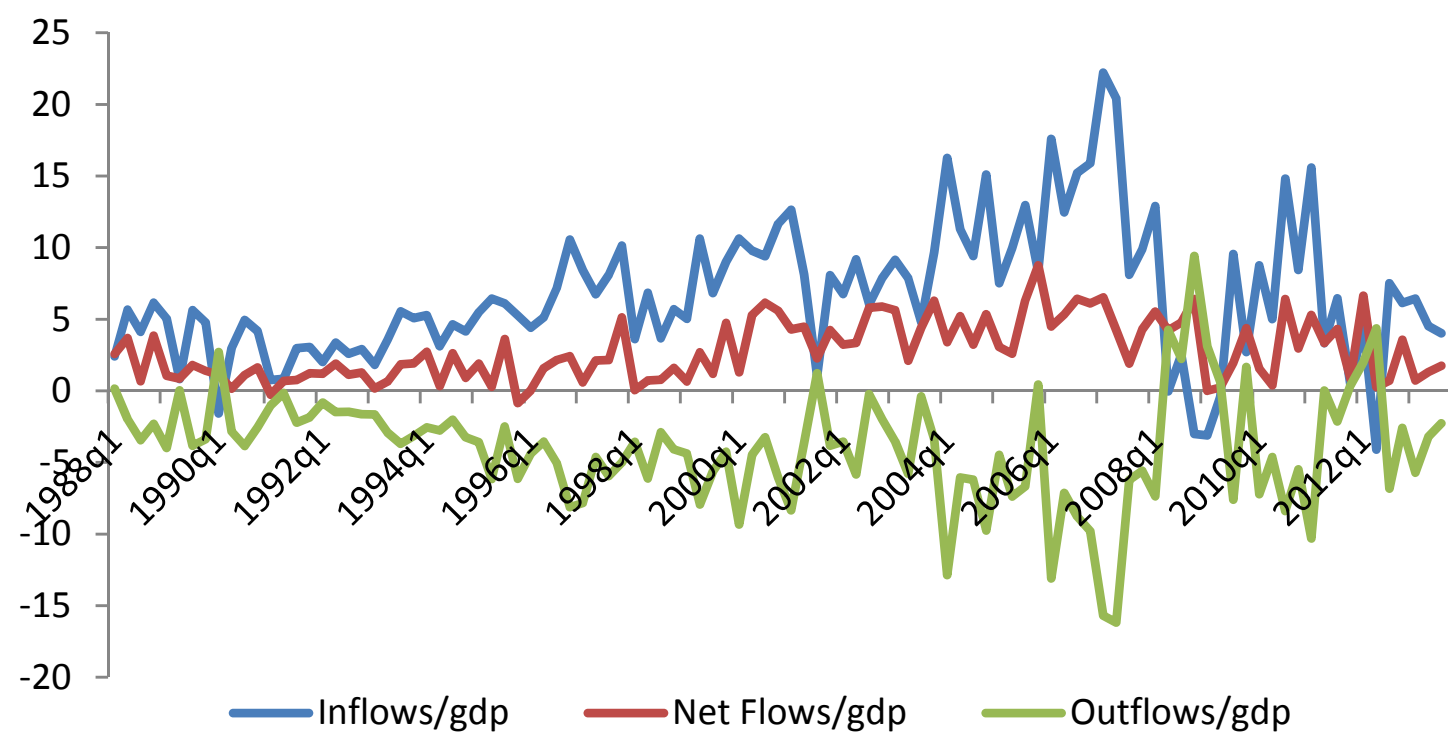

*Notes*: All variables are divided by the GDP. Sample of observation: 1986:1-2013:3.

Figure 4: US gross capital flows and its aggregates.

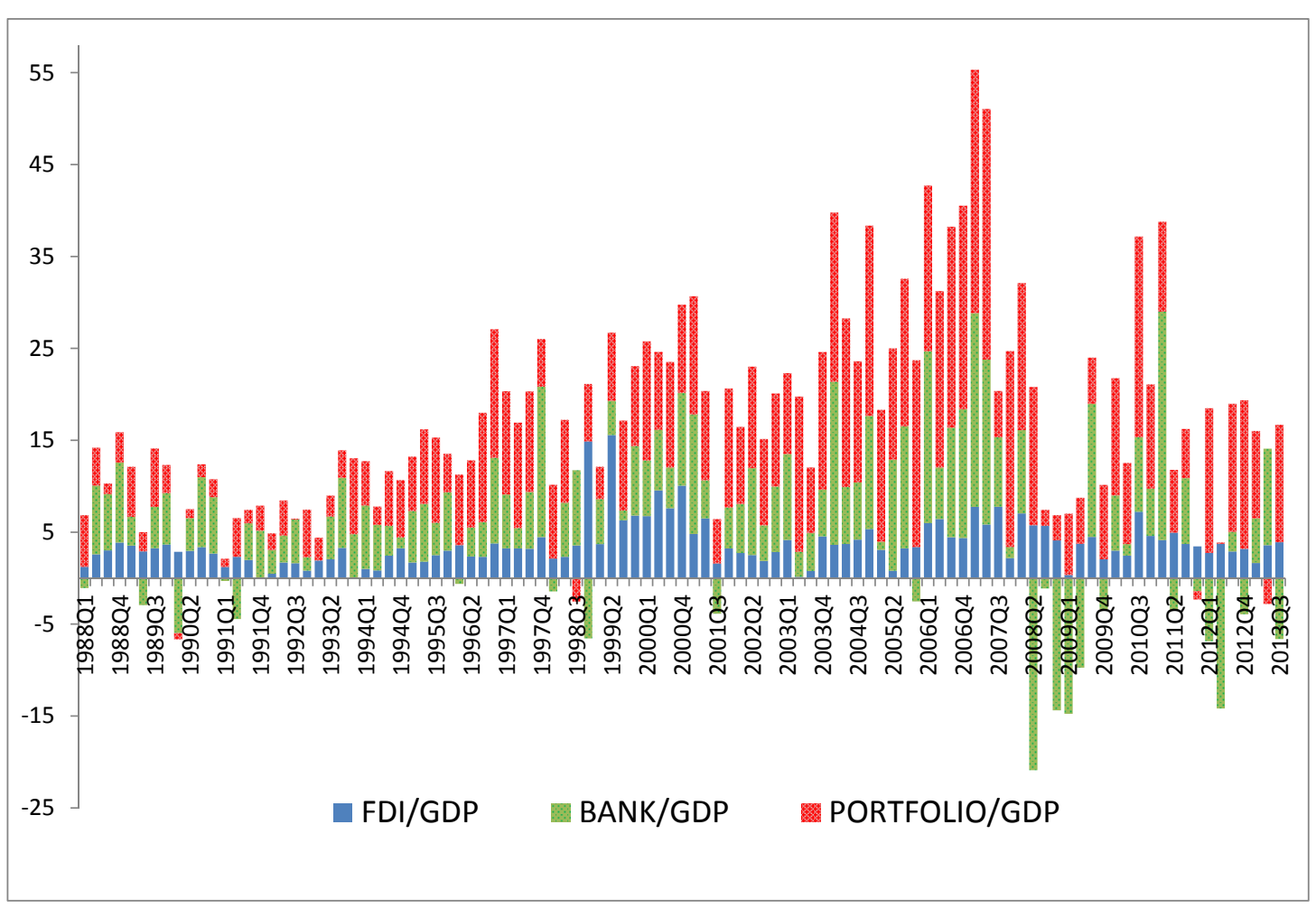

*Notes*: All the aggregates are divided by the GDP. Sample of observation: 1986:1-2013:3.

Figure 5: Response of Capital inflows to monetary policy shocks, volatility shocks and shock to themselves: MIDAS-SVAR model.

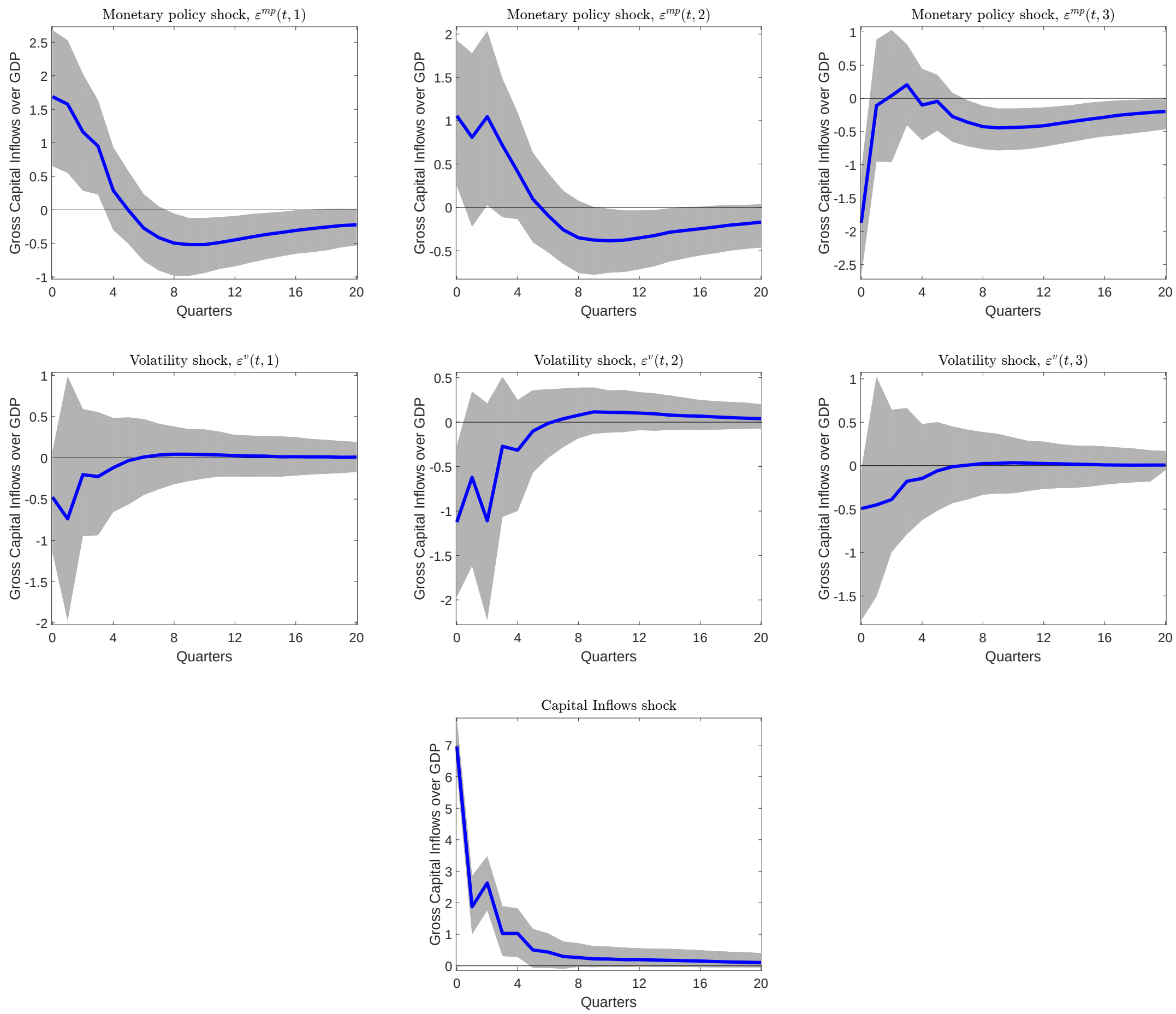

*Notes*: Impulse response functions and 90% bootstrapped confidence intervals for the MIDAS-SVAR model. Response of gross capital inflows over GDP ($k(t)$) to monetary policy shocks (upper panel), volatility shocks (middle panel), and capital shocks (bottom panel). "1", "2" and "3" indicate whether the shocks affect the low frequency $k(t)$ variable during the first, second and third month of the quarter, respectively. The sample period is 1986:1-2013:3.

Figure 6: Response of interest rate to monetary policy shocks: MIDAS-SVAR model.

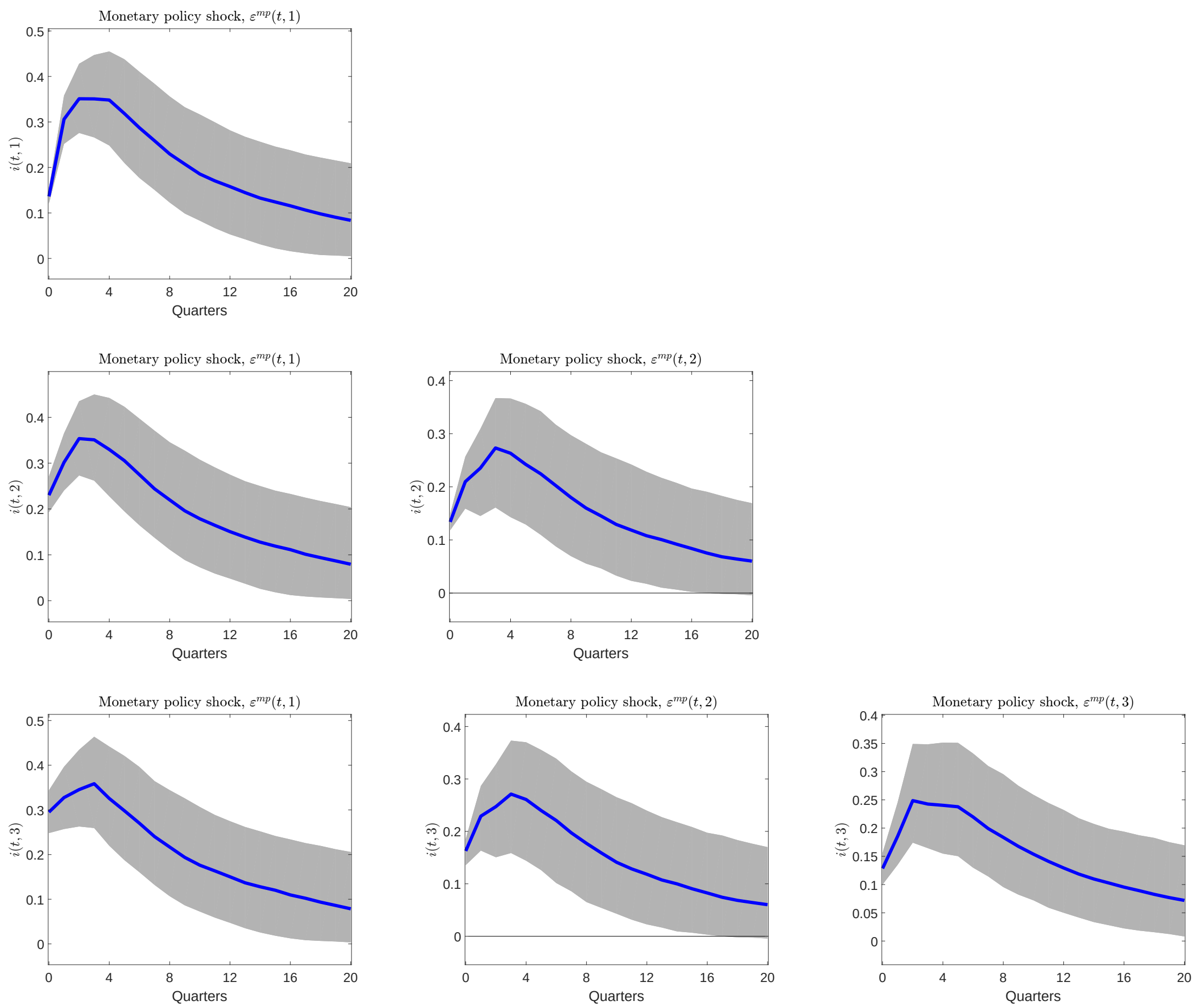

*Notes*: Impulse response functions and 90% bootstrapped confidence intervals for the MIDAS-SVAR model. Response of Fed Funds rates $i(t,1)$ (upper row), $i(t,2)$ (middle row), $i(t,3)$ (lower row) to monetary policy shocks $\varepsilon^{mp}(t,1)$ (left panel), $\varepsilon^{mp}(t,2)$ (middle column) and $\varepsilon^{mp}(t,3)$ (right column). The sample period is 1988:1-2013:3.

Figure 7: Response of VIX to monetary policy shocks: MIDAS-SVAR model.

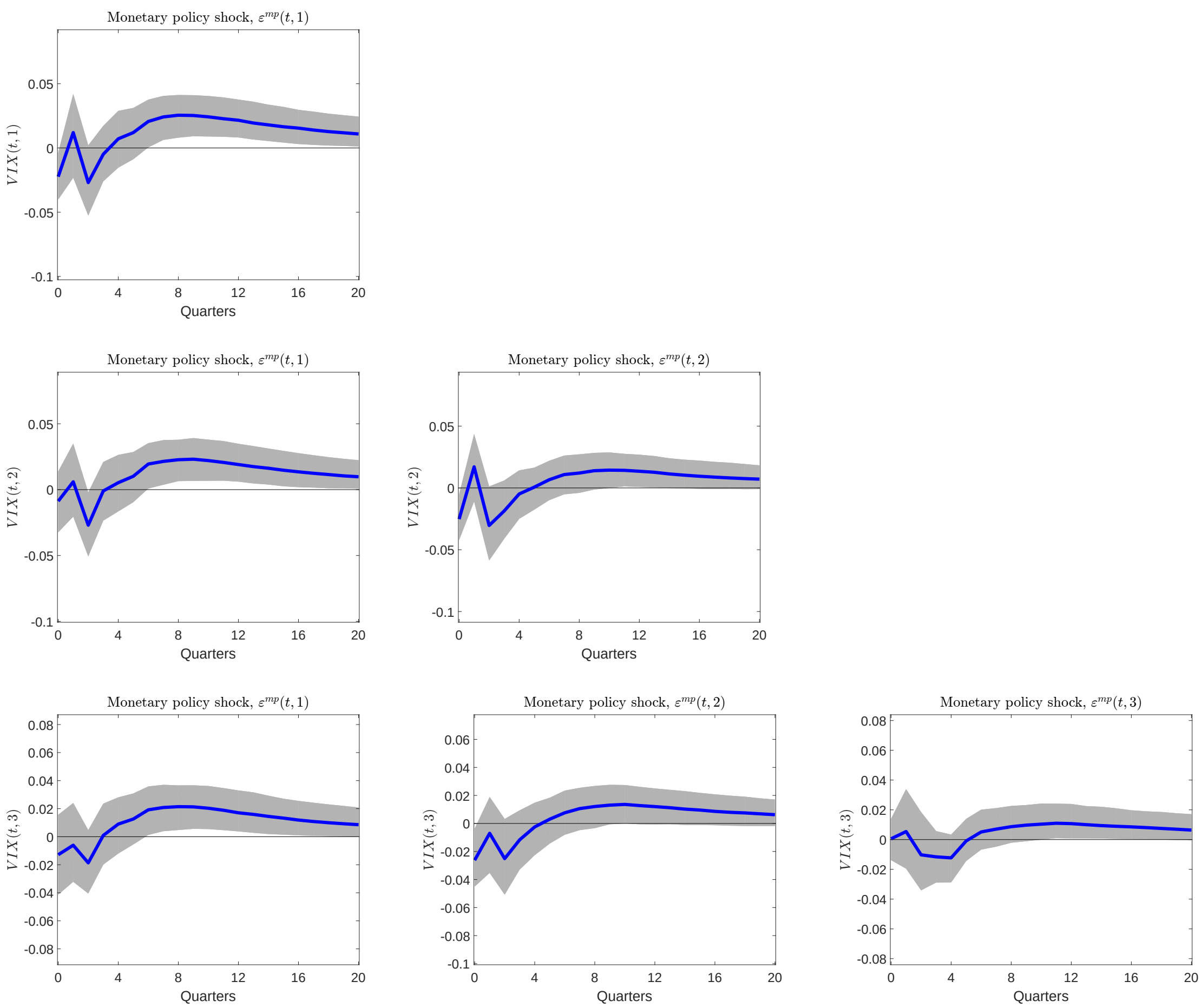

*Notes*: Impulse response functions and 90% bootstrapped confidence intervals for the MIDAS-SVAR model. Response of $vix\,(t,1)$ (upper row), $vix\,(t,2)$ (middle row), $vix\,(t,3)$ (lower row) to monetary policy shocks $\varepsilon^{mp}\,(t,1)$ (left panel), $\varepsilon^{mp}\,(t,2)$ (middle column) and $\varepsilon^{mp}\,(t,3)$ (right column). The sample period is 1988:1-2013:3.

Figure 8: Response of interest rate to volatility shocks: MIDAS-SVAR model.

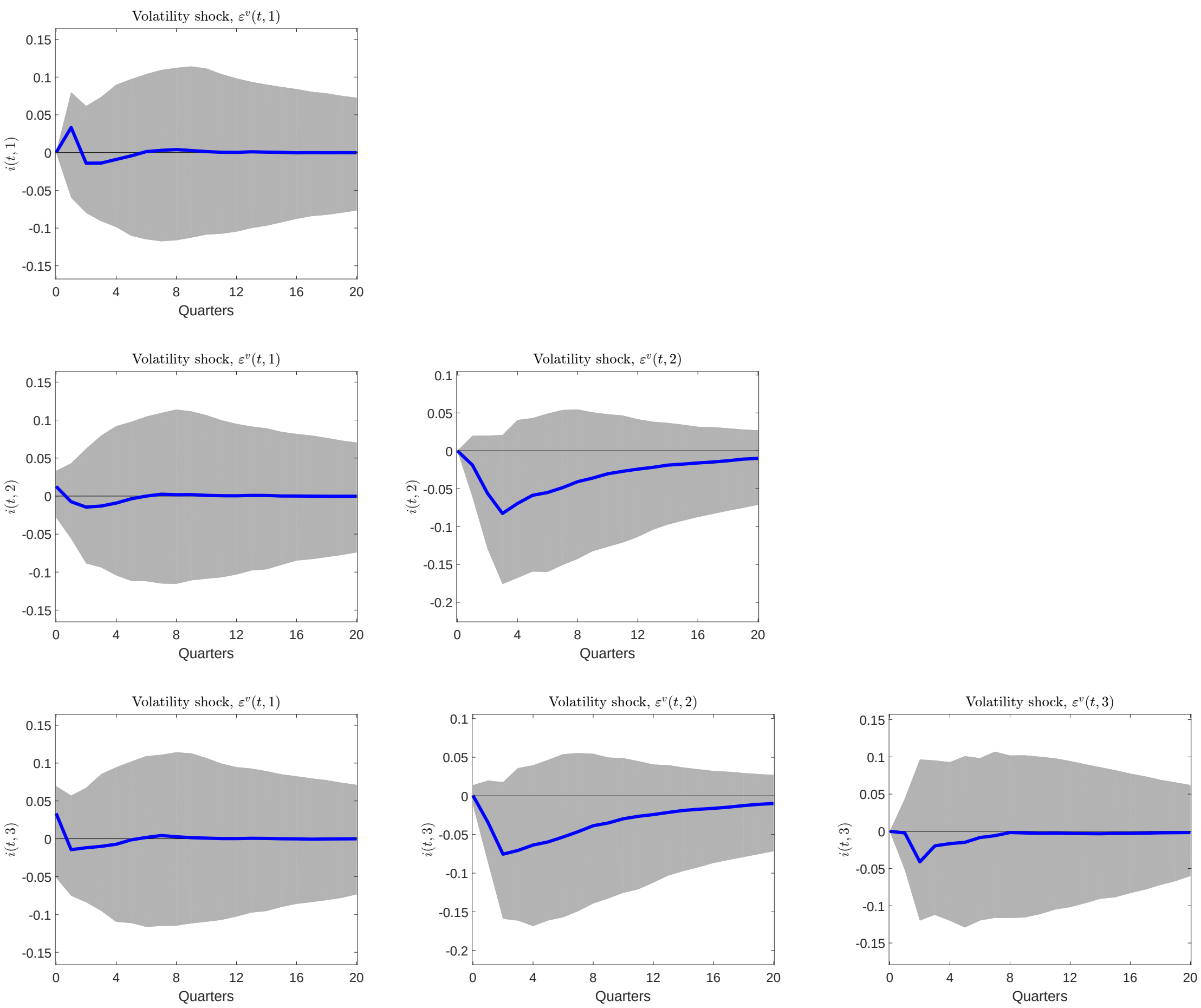

*Notes*: Impulse response functions and 90% bootstrapped confidence intervals for the MIDAS-SVAR model. Response of Fed Funds rates $i\,(t,1)$ (upper row), $i\,(t,2)$ (middle row), $i\,(t,3)$ (lower row) to volatility shocks $\varepsilon^{eu}\,(t,1)$ (left panel), $\varepsilon^{eu}\,(t,2)$ (middle column) and $\varepsilon^{eu}\,(t,3)$ (right column). The sample period is 1988:1-2013:3.